\documentclass[]{elsart}

\usepackage{epsfig}

\usepackage{amssymb}
\usepackage{amsfonts}
\usepackage{latexsym}

\begin{document}

\begin{frontmatter}
\title{Spin, spin-orbit, and electron-electron interactions in mesoscopic systems$^*$}
\author[Weizmann]{Yuval Oreg},
\author[Cornell] {P.~W.~Brouwer},
\author[Cornell] {X. Waintal} and
\author[Harvard] {Bertrand I.~Halperin}
\address[Weizmann] {Department of Condensed Matter Physics,
Weizmann Institute of Science, Rehovot 76100}

\address[Cornell]{Laboratory of Atomic and Solid State Physics,
Cornell University, Ithaca, NY 14853-2501 USA}

\address [Harvard]{ Lyman Laboratory of Physics, Harvard University, Cambridge MA
02138 USA}

\begin{abstract}
 We review recent theoretical developments about the role of spins,
electron-electron interactions, and spin-orbit coupling in metal
nanoparticles and semiconductor quantum dots. For a closed system,
in the absence of spin-orbit coupling or of an external magnetic
field, electron-electron interactions make it possible to have
ground states with spin $S > 1/2$. We review here a theoretical
analysis which makes predictions for the probability of finding
various values of spin $S$ for an irregular particle in the limit
where the number of electron is large but finite. We also present
results for the probability distribution of the spacing between
successive groundstate energies in such a particle. \\
In a metallic particle with strong spin-orbit interactions, for
odd electron number, the groundstate has a Kramers' degeneracy,
which is split linearly by a weak applied magnetic field.  The
splitting may be characterized by an effective $g$-tensor whose
principal axes and eigenvalues vary from one level to another.
Recent calculations have addressed the joint probability
distribution, including the anisotropy, of the eigenvalues. The
peculiar form of the spin-orbit coupling for a two-dimensional
electron system in a GaAs heterostructure or quantum well leads to
a strong suppression of spin-orbit effects when the electrons are
confined in a small quantum dot.  Spin-effects can be enhanced,
however, in the presence of an applied magnetic field parallel to
the layer, which may explain recent observations on fluctuations
in the conductances through such dots.   \\
We also discuss possible explanations for the experimental
observations by Davidovic and Tinkham of a multiplet splitting of
the lowest resonance in the tunneling conductance through a gold
nano-particle.
\end{abstract}
%\begin{keyword}
%\PACS
%\end{keyword}
\end{frontmatter}
\vspace{-0.7cm} {\footnotesize $^*$This article is slated to
appear as a chapter in "Nano-Physics and Bio-Electronics", edited
by T. Chakraborty, F. Peeters, and U. Sivan  (to be published by
Elsevier~Co.).}
\newpage \normalsize
%%%%%%%%%%%%%%%%%%%%%%%%%%%%%%%%%%%%%%%%%%%%%%%%%%%%%%
%% SOME DEFINITIONS FOR MATHEMATICAL SYMBOLS
\def\tensor#1{\protect\ontopof{#1}{\leftrightarrow}{1.15}\mathord{\box2}}
\def\overstar#1{\protect\ontopof{#1}{\ast}{1.15}\mathord{\box2}}
\def\overdots#1{\protect\ontopof{#1}{\cdots}{1.0}\mathord{\box2}}
\def\overcirc#1{\protect\ontopof{#1}{\circ}{1.2}\mathord{\box2}}
\def\loarrow#1{\protect\ontopof{#1}{\leftarrow}{1.15}\mathord{\box2}}
\def\roarrow#1{\protect\ontopof{#1}{\rightarrow}{1.15}\mathord{\box2}}

\def\ontopof#1#2#3{%
{\mathchoice
{\oontopof{#1}{#2}{#3}\displaystyle\scriptstyle}%
{\oontopof{#1}{#2}{#3}\textstyle\scriptstyle}%
{\oontopof{#1}{#2}{#3}\scriptstyle\scriptscriptstyle}%
{\oontopof{#1}{#2}{#3}\scriptscriptstyle\scriptscriptstyle}%
}%
}

\def\oontopof#1#2#3#4#5{%
\setbox0=\hbox{$#4#1$}%
\setbox1=\hbox{$#5#2$}%
\setbox2=\hbox{}\ht2=\ht0 \dp2=\dp0 %
\ifdim\wd0>\wd1 %
\setbox1=\hbox to\wd0{\hss\box1\hss}%
\mathord{\rlap{\raise#3\ht0\box1}\box0}%
\else   %
\setbox1=\hbox to.9\wd1{\hss\box1\hss}%
\setbox0=\hbox to\wd1{\hss$#4\relax#1$\hss}%
\mathord{\rlap{\copy0}\raise#3\ht0\box1}%
\fi
}%
\def\slantfrac#1#2{\kern1em^{#1}\kern-.3em/\kern-.1em_{#2}}
%%%%%%%%%%%%%%%%%%%%%%%%%%%%%%%%%%%%%%%%%%%%%%%%%%%%%%%%%%%%%

\section{ Introduction}
\label{se:intro}

  During the last two decades the fabrication technology of small
conducting islands, known as {\em quantum dots}, using
semiconductor heterostructures, spattering of small metal grains
and other methods have advanced so much that they can behave, under
the right conditions, as {\em artificial atoms} \cite{WF:Review}.
In contrast to natural atoms, these artificial atoms do not have
special symmetries, unless special experimental efforts are being
performed\cite{WF:Review}.

  The electron spin is responsible for a number of interesting
effects in small chaotic conducting islands at low temperatures
which are quite distinct from the role of spin in a bulk material.
The role of spin is modified by electron-electron interactions in
a way that has consequences for the distribution of the energy
separations between ground states with different number of
electrons, as well as for the probability of finding different
spin quantum numbers at a fixed number of electrons. Other
interesting effects are produced by spin-orbit coupling, which is
important distribution of energy levels and wavefunctions in
closed quantum dots and may modifies its properties, and which
affects the conductance distribution for an ensemble of open
quantum dots.
 The addition of source and drain leads, and in case of
semiconductor heterostructures gates that control the charge on
the dot, allows one to measure the properties of the dot-leads
compound as a function of $V$, the difference between the
potentials on the source and  drain,  and $V_{\rm g}$, the
potential on the gate lead.

 In Sec.~\ref{se:interactions} of this paper we discuss a few
effects of electron-electron interactions in a closed dot, with no
spin-orbit coupling and no significant Zeeman field.  The analysis
reviewed in this section has been developed by a number of
research groups in the last few years. We re-derive here an
effective low energy Hamiltonian that was first discussed in
Ref.~\cite{WF:Kurland00} using renormalization group (RG)
arguments. Then, we extend our studies \cite{WF:Brouwer99} for the
ground state spin of a quantum dot, and analyze its influence on
the Coulomb blockade peak spacing that appear in the conductance
as we sweep $V_{\rm g}$ at zero bias voltage, $V$. Further
details, concerning the parameters of the effective model in
actual systems and the relation of the effective model to other
models, are given in the Appendices to the paper.

In Section~\ref{se:SO}, we give a brief review of recent work on
the effects of spin-orbit coupling in metal nanoparticles and GaAs
quantum dots.  The reader will be referred to the literature for a
fuller account.

In Section~\ref{se:mutiplets}, we discuss a problem motivated by
experimental observations of Davidovic and  Tinkham
\cite{WF:Davidovic00} of a multiplet splitting of the {\em lowest}
resonance in the tunneling differential conductance through a gold
nanoparticle as a function of the source-drain potential, $V$. We
consider two possible mechanisms that may lead to effects of that
type. The first is due to an almost degenerate ground state
(Subsection~\ref{se:almost_degenerate}) and the second due to a
nonequilibrium phenomenon (Subsection~\ref{se:nonequilibrium}). In
Subsection~\ref{se:compare} we compare these mechanisms, which
both involve electron spin and/or electron-electron interaction in
an essential way, and suggest experimental ways to distinguish
between them.

\section{Electron-Electron Interactions in a Closed Dot}
\label{se:interactions}

  When the shape of the dot is symmetric \cite{WF:Review}, a part
of its single particle levels are degenerate, just like in an atom
with a spherically symmetric potential. The exact electron many
body configuration, which is set by the repulsive interaction
between the electrons combined with the Pauli exclusion principle,
is summarized in a set of rules, known as Hund's rules. The first
of them assert that a partially filled set of degenerate levels
will have the maximal spin that is consistent with the exclusion
principle. This happens because for a system of several electrons
the ``most antisymmetric'' coordinate wave function has the
largest spin. The most antisymmetric wave function minimizes the
Coulomb repulsion between the electrons because it vanishes when
there are two electrons at the same point. Strong spin-orbit
interactions may change Hund's rules.

To create symmetric dots, special (experimental) effort is
necessary. Any generic dot, however, does not have any special
symmetry and can be considered as a chaotic one, its single
particle levels may assumed to be random, and are described by
random matrix theory (RMT). Many theories have concentrated on the
statistical properties of random levels and the way they can be
used to understand chaotic dots in actual experiments, for review
see Ref.~\cite{Beenakker}. The interactions between electrons in
the dot and between them and the environment were commonly
described by the so-called {\em constant interaction model}. In
this model all the effects of interaction are cast into a single
capacitance that describes the change in the energy of the system
due to the dot's charging.

This class of models was very successful in explaining and
predicting experimental results as Coulomb peak height
fluctuations, conductance fluctuations through an open dot and so
on \cite{Beenakker}. However, it fails to explain experiments
measuring motion of peaks in magnetic field \cite{NT:Oreg00},
distributions of Coulomb blockade peak spacing
\cite{WF:Sivan96,WF:Simmel97,WF:Patel98,WF:Luscher01}, and
multiplets that appear in a single Coulomb blockade peak
\cite{WF:Davidovic00} . Motivated by the failure of the constant
interaction model, a few models that extend it were suggested
\cite{WF:Kurland00,WF:Brouwer99,WF:Baranger00,WF:Ullmo01,WF:Aleiner01}.
In Subsection~\ref{se:Heff} we derive, by integrating out fast
degrees of freedom in the RG sense,  an effective
low-energy-Hamiltonian, $\mathcal{ H}_{\rm eff}$. This effective
Hamiltonian, first discussed in Ref.~\cite{WF:Kurland00},
describes the properties of the quantum dot at energies smaller
than the Thouless energy $E_{\rm Th}$ of the system, which is
inversely proportional to the time it takes to cross the dot along
its largest dimension. We then show that a proper choice of
parameters for $\mathcal{H}_{\rm eff}$ and its analysis reproduced
other models for the effects of electron--electron interaction in
quantum dots. Appendix~\ref{ap:toy} presents some details of the
relation and equivalence of different models. Throughout this
article we will neglect fluctuations of the interaction
parameters, we discuss the limits of this approximation in
Subsection \ref{se:fluctuations}.

 In the process of averaging over the fast motion of
the electrons in the Fermi sea, three channels of interaction
appear: The direct/charging channel, the exchange/spin channel and
the Cooper channel. The actual values of the parameters that
describe these channels in $\mathcal{H}_{\rm eff}$ depend on the
fast motion of electrons. We can have a richer situation when
there are intermediate scales between the Fermi energy and the
Thouless energy, {\em e.g.}: the inverse of the mean free time,
$\tau$, between elastic collisions with static impurities in
diffusive systems. In Appendix \ref{ap:LP} we find, using the
Fermi-liquid-theory, the effective exchange interaction parameter
for several three and two dimensional systems. Its dependance on
electron density in two dimensional Si-MOSFET and AlGaAs ballistic
heterostructures is estimated. Appendix~\ref{ap:cooper} deals with
the renormalization of the Cooper channel.

After a detailed discussion of the model in
Subsection~\ref{se:Heff}, we use it to analyze several physical
quantities. In Subsection~\ref{sec:GroundStateSpinDistribution} we
present an extended version of our study in
Ref.~\cite{WF:Brouwer99} of the ground state spin configuration
distribution. The following Subsection (Subsection~\ref{se:CB})
discusses the effects of spin fluctuations in the ground state on
the Coulomb blockade peak distribution, (see also
Ref.~\cite{WF:Ullmo01}). In Subsection~\ref{se:exp} we compare
some predictions of the theory with available experimental results
\cite{WF:Patel98,WF:Luscher01}. We find that, although several
features of the theory are observed in experiments, there is still
disagreement between theory and experiment. We will discuss the
possible causes for this discrepancy.

\subsection{Effective Hamiltonian}
\label{se:Heff}

 To find properties, such as the ground state spin distribution of
electrons in the dot in the presence of electron--electron
interactions, we would like to describe them at low energy in a
simple form. Ref.~\cite{WF:Kurland00} shows that in the limit of
large dots the effective Hamiltonian, at energies smaller than
$E_{\rm Th}$, is:
\begin{equation}
  {\mathcal H}_{\rm eff} = \sum_{\mu,s} \varepsilon_{\mu s}  \psi^{\dagger}_{\mu s} \psi_{\mu s}
 + J_{\rm s} \vec{S} \cdot \vec{S} + J_{\rm c}  T^{\dagger}  T + U_{\rm c}  N^2. \label{eq:heff}
\end{equation}
Here, $ N=\sum_{\mu, s = \uparrow, \downarrow}  \psi^\dagger_{\mu
s} \psi_{\mu s}$, $ \vec {S} = \sum_{\mu, s, s'} \frac{1}{2}
\psi^\dagger_{\mu s} {\vec {\sigma}}_{s,s'}  \psi^{\hphantom
\dagger}_{\mu s'}$ and $ T = \sum_\mu  \psi^\dagger_{\mu \uparrow}
\psi^\dagger_{\mu \downarrow}$. The $\mu$ sum runs over $M \equiv
g=2\pi E_{\rm Th}/\Delta$ states. The symbol $\Delta= 1/(\nu
\mathcal{V})$ denotes the average single particle level spacing in
a dot of volume $\mathcal{V}$ and thermodynamic density of states
$\nu$ (per spin). Notice that in our notation $\Delta$ refers to
the level spacing for a single spin state in the dot. In a dirty
dot of length $L$ and diffusive constant $D$, $E_{\rm Th}=D/L^2$,
while for a single dot we replace $D$ by $\sim v_F L$.

The first term in (\ref{eq:heff}) is universal: $\varepsilon_\mu$
is a single electron  eigenenergy of a random matrix, and the
operator $\psi^\dagger_{\mu,s}$ is the creation operator of an
electron with spin $S$ at an eigenstate $\mu$ of a random matrix.
The set $\{ \varepsilon_\mu \}$ depends on the symmetry of the
problem. In the presence of time reversal symmetry the
eigenenergies are taken from the Gaussian orthogonal ensemble
(GOE, $\beta=1$), and in the absence of time reversal symmetry
from the Gaussian unitary ensemble (GUE, $\beta=2$). When strong
spin-orbit interactions are present they are taken from the
Gaussian symplectic ensemble (GSE, $\beta=4$)\footnote{ The
symmetry index $\beta$ counts the degrees of freedom of the matrix
elements of the single particle Hamiltonian, $\beta=1,2,\mbox{ or
} 4$ if its elements are real complex or real quaternion numbers,
respectively.  A magnetic flux $\sim \phi_0 /\sqrt{g}$ through the
dot, where $\phi_0=h c/e \approx 4.12 \times 10^{-13} Tesla \cdot
m^2$, leads to a time reversal symmetry breaking.}.

The direct Coulomb interaction constant $U_{\rm c}$, the exchange
constant $J_{\rm s}$, and the interaction in the Cooper channel
$J_{\rm c}$ depend on the specific system and the model one uses
for the interaction. In addition, there are non universal
corrections to Hamiltonian (\ref{eq:heff}) that vanish, however,
in the limit of $g \rightarrow \infty$. (See also in Subsection
\ref{se:fluctuations}.) When the time reversal symmetry is broken
the interaction in the Cooper channel vanishes. We will see below
that even in the presence of time reversal symmetry the
interaction in the Cooper channel is reduced due to a ``screening"
by fast electrons(see also Appendix \ref{ap:cooper}).

To understand what are the effective low energy interaction
constants $U_{\rm c}$, $J_{\rm s}$, and $J_{\rm c}$ it is useful
to describe the derivation of the effective Hamiltonian
(\ref{eq:heff}) in terms of a RG scheme. When the temperature
decreases we integrate out progressively the fast motion of the
electrons with energy far from the Fermi energy and find effective
coupling constants in the direct ($U_{\rm c}$), exchange ($J_{\rm
s}$) and Cooper ($J_{\rm c}$) channels
\cite{RFS:Shankar94,RFS:Polchinski93}. First, the fast motion of
the electrons, at energies of $O(E_{\rm F})$ away from the Fermi
level, ``dresses'' the bare electrons and forms quasi-particles.
The Landau Fermi-liquid theory describes this dressing
process\cite{RFS:Shankar94,RFS:Polchinski93} and the way it
renormalizes the system parameters. In Appendix~\ref{ap:LP} we use
this theory to estimate $J_{\rm s}$ for several situations.

 The Fermi-liquid ``dressing'' continues up to energies of order
$1/\tau$. Below $1/\tau$ the motion of the electrons becomes
diffusive and new diffusion singularities appear. In a situation
where one of the dimensions of the system is much smaller than the
others, the system may become quasi-two-dimensional at frequencies
smaller than the Thouless energy that is related to the short
dimension. This reduction in the dimension of the system enhances
the diffusive singularities and changes the flow of the
interaction parameters\cite{DS:Finkelstein90}. It appears that the
RG flow in the Cooper channel is sensitive to disorder more than
the RG flow in the other channels\cite{DS:Finkelstein90}, (see
also Appendix~\ref{ap:cooper}). However, in certain situations,
especially when disorder is strong, we also have to consider its
effect on the flow in the other channels\cite{DS:Finkelstein90}.

Finally, we arrive at temperatures  $T< E_{\rm Th}$ and are left
with an effective ``zero dimensional'' Hamiltonian,
$\mathcal{H}_{\rm eff}$. The length scale associated with such low
temperatures is larger than the system size and therefore the
interaction parameters are constants that do not depend on the
site or state index. We note that in ballistic samples, {\em
i.e.}, when the electrons cross the sample before suffering
substantial scattering from impurities we may use the Fermi-liquid
theory, without additional complications due to the diffusive
motion at intermediate scales.
\subsubsection{Fluctuations in the interaction parameters}
\label{se:fluctuations} \label{pg:flactuations}
 In practice, not all the samples have exactly the same shape
and/or impurity configuration, we therefore expect to find sample
to sample (mesoscopic) fluctuations in the RG process that will
lead to fluctuations in the interaction constants of the different
channel. By assumption (that is motivated and supported by
numerical analysis \cite{WF:Berkovits98,WF:Levit99}) the sample to
sample fluctuations in the single particle levels are described by
random matrix theory. The eigenenergies and eigenfunctions are
those of a random matrix.

The random electron states have random charge distributions and
hence we expect that the interaction parameters $J_{\rm c}$,
$U_{\rm c}$ and $J_{\rm s}$ themselves will fluctuate with the
electron number. The interaction in the Cooper channel is reduced
by a logarithmic factor (see Appendix~\ref{ap:cooper}) and we will
neglect its fluctuations.

When an electron is added to the system charge flows to the dot
edges and leads to fluctuations in the self-consistent nonuniform
potential \cite{WF:Blanter97}. These fluctuations scale as $\sim
(r_{\rm s}/\sqrt{g}) \Delta$, and give the largest contributions
to the fluctuations in $U_{\rm c}$. [See Eqs.~(\ref{def:rs3}) and
(\ref{def:rs2}) for a precise definition of $r_{\rm s}$ in two and
three dimensions.] Fluctuations in $U_{\rm c}$ are relevant to the
Coulomb peak spacing distribution (Subsection~\ref{se:exp}) and to
nonequilibrium effects (Subsection~\ref{se:nonequilibrium}) in the
conductance through the dot at finite bias voltage.

The short range part of the Coulomb interaction determines the
exchange integral in the expression for the exchange interaction
parameter $J_{\rm s}$.  We can therefore use a contact interaction
model  to find its fluctuations. The fluctuations in the
interaction parameter in that case, for three dimensional samples,
are~\cite{WF:Blanter96} $\sqrt{\mbox{\rm var}\;{J_{\rm s}}}
\approx J_{\rm s} \max\left\{ 1/(2 \sqrt{2 N}), c_3/g\right\}, $
where $c_3$ is a numerical number of order $1$, and $N \sim
E_F/\Delta$ is the total number of electrons in the dot. The first
term is larger for $g^2\gg N \Leftrightarrow L < (k_F l) l$, with
$L$ the sample size, $l$ the mean free path, and $k_F$ the Fermi
momentum. For ballistic systems $l \approx L$, and the first term
is much larger. Semiclassically~\cite{WF:Blanter98}, the first
contribution arises from direct trajectories between two points in
the dot, and  the second is built from indirect trajectories with
possible scattering on the surface or on static impurities. In two
dimensional samples a similar calculation
\cite{WF:Ullmo01,WF:Aleiner01} gives, $\sqrt{\mbox{var}\;{J_{\rm
s}}} \approx J_{\rm s} \max\left\{c_2 \log{N}/ \sqrt{N},
c_2'/g\right\}, $ where $c_2$ and $c_2'$ are of order $1$. For
ballistic systems we find that, as in three dimensional samples,
the first term is larger.

The actual contribution of the fluctuations in $J_{\rm s}$ to the
fluctuations in the ground state energy is larger by a factor
$\sim \sqrt g$,  as in the calculation of the ground state energy
we have to include interaction of a single spin with $\sim g$
electrons~\cite{WF:Ullmo01}.

In any case, in the universal limit, when $g, N \rightarrow
\infty$ we can neglect the fluctuations in the interaction
parameters.

 \subsubsection{A toy model with contact interaction}

In Ref.~\cite{WF:Brouwer99} we discussed a model with contact
interaction, described by the Hamiltonian
\begin{equation} \label{eq:ham}
  {\mathcal H_{\rm toy}} = \sum_{n,m,s} c^{\dagger}_{n,s}
{\mathcal H}_0(n,m) c^{\vphantom{\dagger}}_{m,s} +
      u M \sum_{n} c^{\dagger}_{n,\uparrow} c^{\dagger}_{n,\downarrow}
      c^{\vphantom{\dagger}}_{n,\downarrow}
      c^{\vphantom{\dagger}}_{n,\uparrow}.
\end{equation}
Now $m$ runs over $M$ coarse grain sites (in real space) and
${\mathcal H}_0(n,m)$ is a random matrix. The parameter $u$
describes the strength of the interaction between the electrons.

This toy model was analyzed within the self consistent
Hartree--Fock approximation in the limit of large $M$
\cite{WF:Brouwer99}.
 We will show in Appendix~\ref{ap:toy} that to first order in $u$ the toy
model (\ref{eq:ham}) is equivalent to the Hamiltonian
(\ref{eq:heff}) with parameters
\begin{equation}
\label{eq:param}
 U_{\rm c}= u/4,\;\; J_{\rm s} = -u \mbox{ and } J_{\rm c}=u.
\end{equation}
Higher order corrections in $u$ preserve the symmetry and the
structure of Hamiltonian~(\ref{eq:heff}) but give different values
for its parameters. In particular, as shown in
Appendix~\ref{ap:cooper} for positive $u$ (which corresponds to a
repulsive interaction) they reduce $J_{\rm c}$ by a factor
$\propto \log M$.

\subsection{Ground state spin distribution}
\label{sec:GroundStateSpinDistribution}
\begin{figure}[ht]
    \vglue 0cm \epsfxsize= 0.8\hsize
    \hspace{0.1\hsize}
    \epsffile{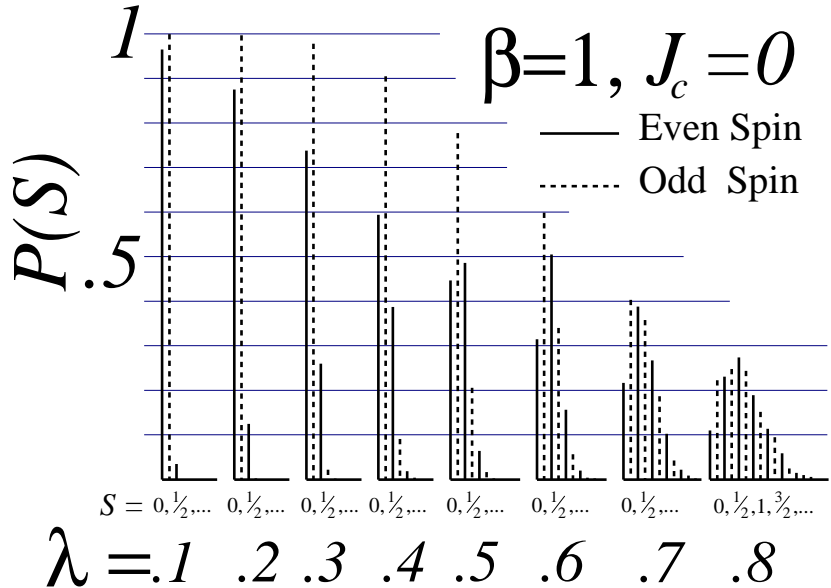}
    \vspace{-0.3cm}
\caption{\label{fg:Ps1} The probability distribution $P(S)$ of the
  ground state spin of a quantum dot, computed from
  Eq.~(\protect\ref{eq:EGdiff}) for different values of the
  interaction parameter $\lambda = -J_{\rm s}/\Delta$ with time
  reversal symmetry. Solid histograms are for integer spins, dotted
  ones for half-integer spins. All the graphs are starting with
  spin 0 increasing to the right in increments of 1/2 spin.}
\end{figure}
In this section we will study the ground state spin of Hamiltonian
(\ref{eq:heff}) assuming that the interaction constant $J_{\rm
c}=0$ [as it renormalized towards zero by electrons with energy
larger than the Thouless energy and is further renormalized within
the toy model (see Appendix~\ref{ap:cooper})]. In this case, the
Hamiltonian (\ref{eq:heff}) becomes noninteracting within each
spin sector. For a fixed number of particles $N$ we can neglect
also the charging energy $U_{\rm c}$. The spin of the ground state
is then found by minimizing the energy $E_{\rm G}(S)$ of the
lowest lying state with total spin $S$, as a function of $S$.
Since the lowest energy state with spin $S$ has precisely $2S$
singly occupied states, all lower lying states being doubly
occupied, one has~\cite{WF:Brouwer99}
\begin{eqnarray}
  E_{\rm G}(S) - E_{\rm G}(S_0) &=&
  \sum_{\mu=1}^{S-S_0} (\varepsilon_{N+\mu+2 S_0} - \varepsilon_{N+1-\mu}) \nonumber \\
  && \mbox{} + J_{\rm s} \left[ S(S+1) - S_0(S_0+1) \right].
  \label{eq:EGdiff}
\end{eqnarray}
where $S_0 = 0$ ($1/2$) if the total number of particles $N$ is
even (odd).
\begin{figure}[ht]
 \vglue 0cm \epsfxsize=0.8\hsize
  \hspace{0.1\hsize}
  \epsffile{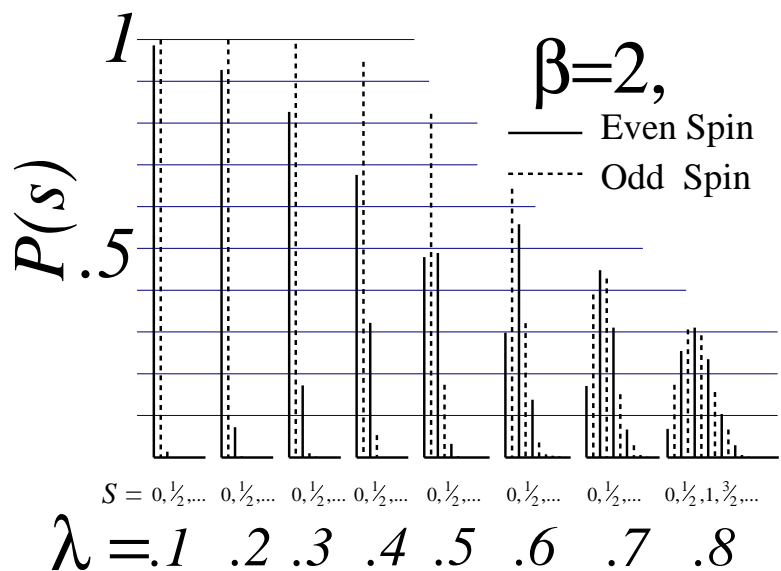}\vspace{-1.4cm}
\caption{\label{fg:Ps2} The probability distribution $P(S)$ of the
  ground state spin of a quantum dot, computed from
  Eq.~(\protect\ref{eq:EGdiff}) for different values of the
  interaction parameter $\lambda$ without time reversal symmetry.
  Solid histograms are for integer spins, dotted ones for
  half-integer spins. All the graphs are starting with spin 0
  increasing to the right in increments of 1/2 spin.}
\end{figure}
Since the precise positions of the energy levels
$\left\{\varepsilon_{\mu} \right\}$ in Eq.~(\ref{eq:EGdiff})
fluctuate from grain to grain, the ground state spin $S$ does so
as well. Plots of the probability distribution $P(S)$ are shown in
Fig.~\ref{fg:Ps1} and Fig.~\ref{fg:Ps2} for different values of
the dimensionless parameter $\lambda = -J_{\rm s}/\Delta$. The
effective parameter $\lambda$, that includes renormalization from
fast electrons, is calculated for small metallic dots and
semiconductors dots in Appendix~\ref{ap:LP}. The distributions are
obtained by taking the levels $\varepsilon_{\mu}$ from the GOE
($\beta=1$), or, when time-reversal symmetry is broken, from the
GUE, ($\beta=2$), and minimizing Eq.~(\ref{eq:EGdiff}) with
respect to~$S$.
The way spin-orbit coupling reduces the effect of the interaction
in the exchange channel is discussed in Sec.~\ref{se:so+int}.

\subsection{Application to Coulomb Blockade statistics}
\label{se:CB}

A few experimental results on the statistics of the Coulomb
blockade peak spacing in a quantum dot
\cite{WF:Sivan96,WF:Simmel97,WF:Patel98} suggest that the
predictions of the constant interaction model, with
single-particle levels taken from RMT and with a constant charging
interaction $U_{\rm c}$ only, fails to describe the fluctuations
of Coulomb blockade peak spacing. The RMT+$U_{\rm c}$-model
predicts a Wigner surmise peak distribution, and even-odd effects.
But experimentally, the peak spacing distribution is roughly
Gaussian (with non Gaussian tails), it width is $\gtrsim \Delta$
\cite{WF:Patel98}, and even-odd effects are not observed.
Numerical studies for a few electrons, with mutual Coulomb
repulsion, in a disordered medium~\cite{WF:Berkovits98} deviate
from the predictions of the RMT+$U_{\rm c}$-model as well.

We will discuss now what the model (\ref{eq:heff}) predicts for
the Coulomb blockade peak spacing distribution. By definition, the
spacing between the $N$'th Coulomb peak and the $N-1$'th Coulomb
peak is given by
$$
\Delta E=(E_{N+1}-E_N)-(E_{N}-E_{N-1})
$$
where $E_N$ is the energy of the system with $N$ electrons.
%There are a few sources of fluctuations in $\Delta E$ as $N$ is changed.
Fluctuations in the single particle energy levels $\{
\varepsilon_\mu \}$, and in the interaction parameters $U_{\rm
c}$, $J_{\rm s}$ and $J_{\rm c}$ may lead to  fluctuations in
$\Delta E$ as $N$ changes. The latter, however, vanish when $g
\rightarrow \infty$. As we discussed in the previous section, the
fluctuations of the single particle levels induce fluctuations in
the ground state spin of the dot even when the interaction
constants $J_{\rm s}$, $U_{\rm c}$ and $J_{\rm c}$ do not
fluctuate with $N$. These, in turn, cause fluctuations in the peak
spacing.

\begin{figure}[ht]
    \vglue 0cm \epsfxsize=0.8\hsize
    \hspace{0.1\hsize}
    \epsffile{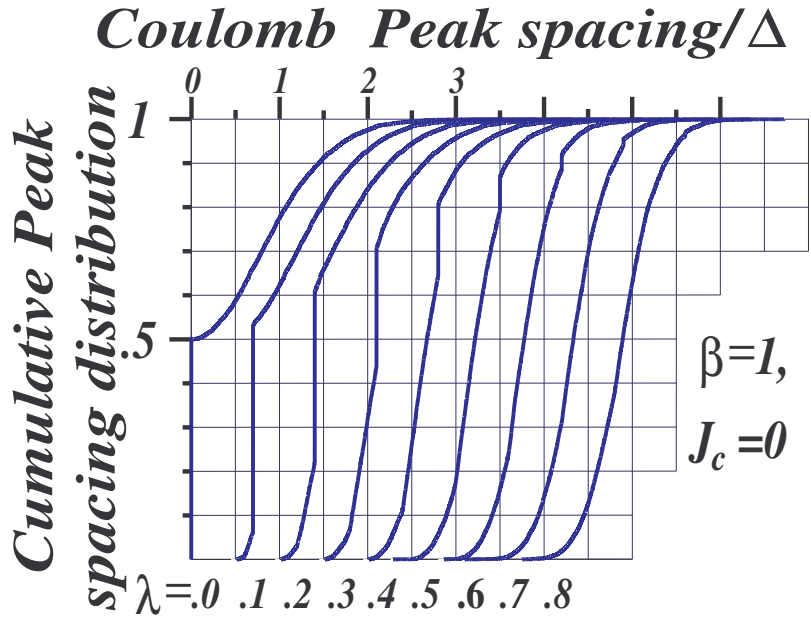}\vspace{-1.5cm}
\caption{\label{fg:b0the}
  Theory for cumulative Peak spacing distribution in the presence of time
  reversal symmetry ($\beta=1$). We show the case without interaction in the Cooper channel
  ($J_{\rm c}=0$) as it is suppressed (see Appendix~\ref{ap:cooper}). For
  clarity we shift the distributions with exchange parameters
  $\lambda=-J_{\rm s}/\Delta$ in intervals of half of the average
  level spacing, $\Delta$.}
\end{figure}
In Fig.~\ref{fg:b0the} we plot the cumulative Coulomb peak spacing
distribution for ensembles with a GOE symmetry. As discussed in
Appendix~\ref{ap:cooper} the interaction in the Cooper channel is
suppressed (by a logarithmic factor). We therefore plot in
Fig.~\ref{fg:b0the} the cumulative peak spacing distribution for a
GOE ensemble without interaction in the Cooper channel($J_{\rm
c}=0$). Fig.~\ref{fg:b0the} shows the distributions for exchange
interaction strengths of $\lambda=-J_{\rm s}/\Delta=0,0.1, \dots
0.8$. We choose to plot the {\em cumulative} peak spacing
distributions and not histograms of the peak spacing density
distribution. Plotting the cumulative distribution allows us to
represent delta functions in the spacing distribution and avoids
the need to work with arbitrarily binning intervals, as is needed
for a histogram.

Examining Fig.~\ref{fg:b0the}, one notices that there is a jump in
the cumulative Coulomb peak spacing distribution at energy $\Delta
E = 2 \lambda \Delta$, corresponding to a delta function in the
spacing distribution at that energy. This occurs because there is
a finite probability that starting from a spin-singlet ground
state, two successive electrons will enter with opposite spin into
the same single-particle state, and the quantity $2 \lambda
\Delta$ is the exchange energy cost for adding the second
electron, when $J_c =0$.
However, at large values of $\lambda$ the probability that the
ground state of the dot is a singlet is smaller, hence the hight
of the jump decreases. There is an additional substructure of the
distributions ({\it e.g.} a kink in the lower part of the
distribution) that becomes smoother when $\lambda$ increases. To
understand a few details of the Coulomb peak spacing distribution
curve, we have to find the ground state energies of a disordered
system with $N-1$, $N$ and $N+1$ electrons. (We assume that the
system has the same disorder realization for consecutive electron
entries.)

\begin{figure}[ht]
    \vglue 0cm \epsfxsize=0.8\hsize
    \hspace{0.1\hsize}
    \epsffile{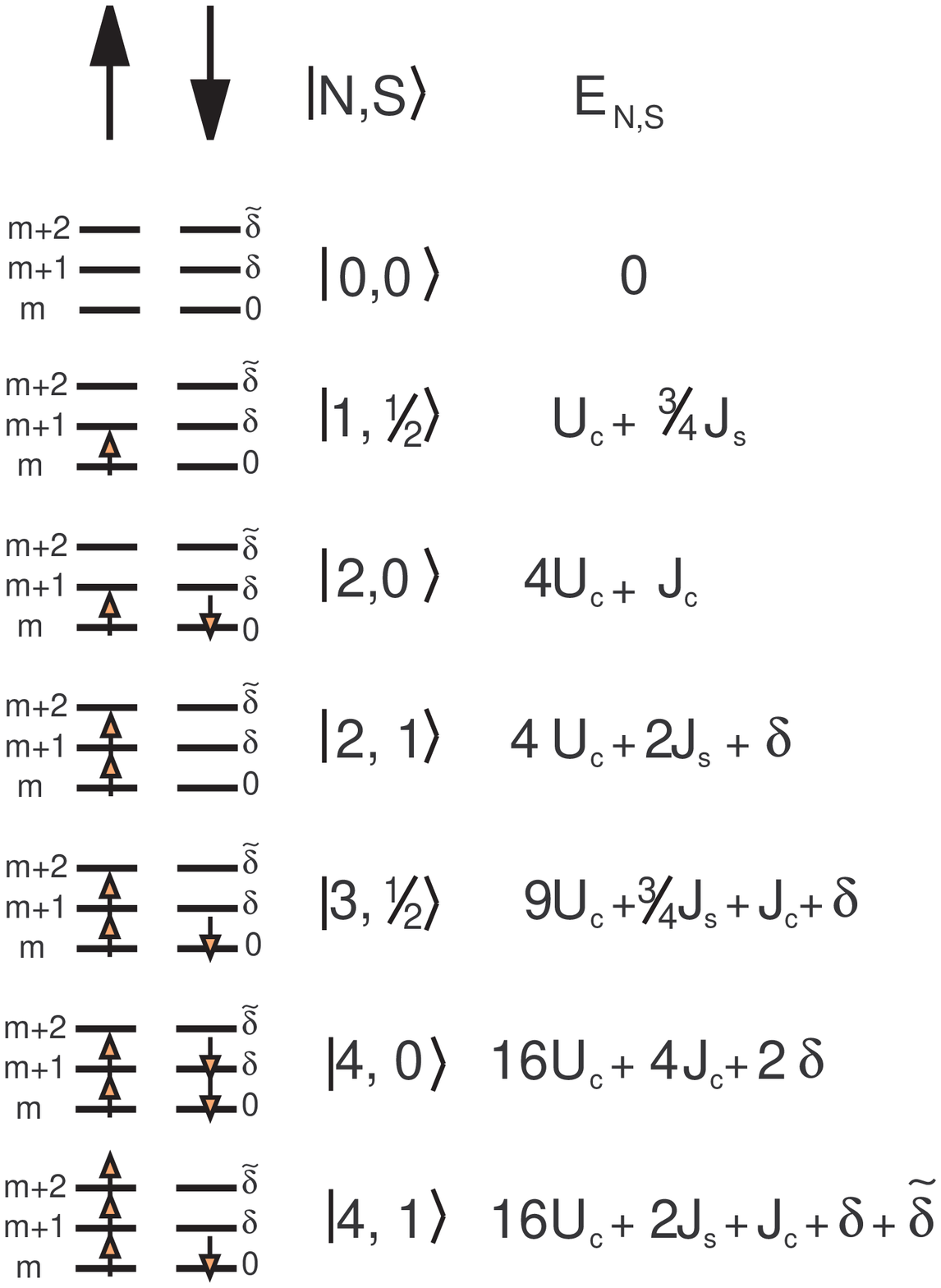}\vspace{0cm}
\caption{\label{fg:spinconf} Possible spin configurations of the
dot and their energies according to Eq.~(\ref{eq:heff}). We assume
that all states below level $m$ are doubly occupied and denote
this many body state $\left| 0,0 \right\rangle$. The single
particle states are: $\epsilon_m=0,\; \epsilon_{m+1}=\delta,\;
\epsilon_{m+2}=\tilde \delta$.}
\end{figure}

The ground state of a system with $N$ electrons is also
characterized by its spin~$S$. Few examples of the energies of
states $\left| N,S \right\rangle$ are summarized in
Fig.~\ref{fg:spinconf}. The Coulomb peak spacing is given by
$\Delta E= E_{N-1, S_{N-1}}+ E_{N+1,S_{N+1}}-2 E_{N, S_N}$, where
$E_{N,S}=\left\langle N, S \right| \mathcal{H}_{\rm eff} \left| N,
S \right\rangle$ is the energy of the state $\left| N,S
\right\rangle$. [Notice that in the presence of interaction in the
Cooper channel $\left| N, S \right\rangle$ is not necessarily an
eigen state of $\mathcal{H}_{\rm eff}$, defined in
Eq.~(\ref{eq:heff})]. Different values of $S_{N-1},S_{N},S_{N+1}$
give rise to different spin sequences.

 There are few possibilities for sequences of spin entries, (see
Table~\ref{tb:peakspacing}). Sequence $\# 1$ describes a situation
where initially there are $(2m-1)$ electrons in the system. The
first $2(m-1)$ single particle states are doubly occupied, and the
last state (state $m$) is singly occupied. We denote this state by
$\left|1, {^1\!/_2} \right\rangle$, see Fig.~\ref{fg:spinconf}.
Then an electron is added to state $m$ so that it is doubly
occupied, we denote this state by $\left|2,0\right\rangle$. The
third electron is added to state $m+1$ so that it singly occupied
to form the state $\left| 3, {^1\!/_2} \right\rangle$. The peak
spacing of this sequence~is:
\begin{eqnarray}
\nonumber \Delta E_{1} &\equiv& E_{1,^1\!/_2}+E_{3, {^1\!/_2}}-2 E_{2,0} \\
\nonumber &=& U_{\rm c}+ {^3\!/_4} J_{\rm s}\;+ \;9 U_{\rm
c}+\delta + {^3\!/_4} J_{\rm s}+J_{\rm c}\;-\; 2( 4 U_{\rm c}+J_{\rm c})\\
\nonumber &=& 2 U_{\rm c} + {^3\!/_2} J_{\rm s} - J_{\rm c}
+\delta = \delta - u - J_{\rm c} \equiv \Delta T_{1}.
\end{eqnarray}
  In the last equality we have used relation (\ref{eq:param})
for the contact interaction toy model. The peak spacings for other
sequences, calculated in a similar way, are summarized in
Table~\ref{tb:peakspacing}.
\begin{table}[h]
\hskip 0.8cm
\begin{tabular}{|l|l|l|l|}
 $i$ & Spin Configurations & $\Delta E_i$ & $\Delta T_i$ \\\hline
  1 & $\left| 1,\; {^1\!/_2} \right\rangle \Rightarrow \left| 2,\; 0 \right\rangle \Rightarrow \left| 3,\; {^1\!/_2} \right\rangle$ & $2U_{\rm c}+{^3\!/_2} J_{\rm s} - \;J_{\rm c} +\delta$&$ -u - J_{\rm c} +\delta $\\
  2 & $\left| 1,\; {^1\!/_2} \right\rangle \Rightarrow \left| 2,\; 1 \right\rangle \Rightarrow \left| 3,\; {^1\!/_2} \right\rangle$ & $2U_{\rm c}-{^5\!/_2} J_{\rm s} + \;J_{\rm c} -\delta$&$ 3u + J_{\rm c} -\delta $\\
  3 & $\left| 2,\; 0 \right\rangle \Rightarrow \left| 3,\; {^1\!/_2}\right\rangle \Rightarrow \left| 4,\; 0 \right\rangle$ & $2U_{\rm c}-{^3\!/_2} J_{\rm s} + \;J_{\rm c} \hphantom{+\delta}$&$ 2u + J_{\rm c}\hphantom{-\delta}$\\
  4 & $\left| 2,\; 0 \right\rangle \Rightarrow \left| 3,\; {^1\!/_2}\right\rangle \Rightarrow \left| 4,\; 1 \right\rangle$ & $2U_{\rm c}+{^1\!/_2} J_{\rm s} \hphantom{+2J_{\rm c}} +\tilde \delta$&$ \hphantom{2u + J_{\rm c}} \tilde \delta $\\
  5 & $\left| 2,\; 1 \right\rangle \Rightarrow \left| 3,\; {^1\!/_2}\right\rangle \Rightarrow \left| 4,\; 0 \right\rangle$ & $2U_{\rm c}+{^1\!/_2} J_{\rm s} + 2J_{\rm c} +\delta$&$ \hphantom{2u}+2J_{\rm c} + \delta $\\
  6 & $\left| 2,\; 1 \right\rangle \Rightarrow \left| 3,\; {^1\!/_2}\right\rangle \Rightarrow \left| 4,\; 1 \right\rangle$ & $2U_{\rm c}+{^5\!/_2} J_{\rm s} - \;J_{\rm c} +\tilde{\delta}$&$ -2u -J_{\rm c} + \tilde{\delta} $\\
\end{tabular}
\vskip 0.5cm \caption{\label{tb:peakspacing} \vskip -0.95cm \hskip
2cm The symbol $\Delta E_{i}$ denotes the spaces between the
Coulomb peaks for spin sequences $i$. [$\Delta T_{i}$ is $\Delta
E_{i}$ using the parameters of the toy model,
Eq.~(\ref{eq:param}).] Sequences that involve higher spins are
also possible and are not included in this table. To find the
actual contribution to the peak spacing we should include the
probability that such a sequence occurs.} \vskip -0.5cm
\end{table}

 Sequence $\# 1$ occurs only if $E_{2,0} < E_{2,1} \Leftrightarrow
\delta > J_{\rm c} - 2 J_{\rm s} \Leftrightarrow \Delta T_{1}> u$.
In a similar way one can check that sequence $\# 2 $ occurs when
$\Delta T_2 > u$. Thus, both processes $\# 1$ and $\# 2$ will lead
to a step in the peak spacing distribution at $u$. For the
cumulative peak spacing distribution, this will lead to a
discontinuity in the slope of the curve, which may be seen in
Fig.~\ref{fg:b0the} at the points $\Delta E = \lambda
\Delta=-J_{\rm s}$. Processes $\# 1$ and $\# 2$  are the main
contributions to the approximately linear portions of the curves
in the range $-J_{\rm s}=\lambda \Delta < \Delta E < 2 \lambda
\Delta=-2 J_{\rm s}$, which are seen at small values of $\lambda$.

\begin{figure}[ht]
    \vglue 0cm \epsfxsize=0.8\hsize
    \hspace{0.1\hsize}
    \epsffile{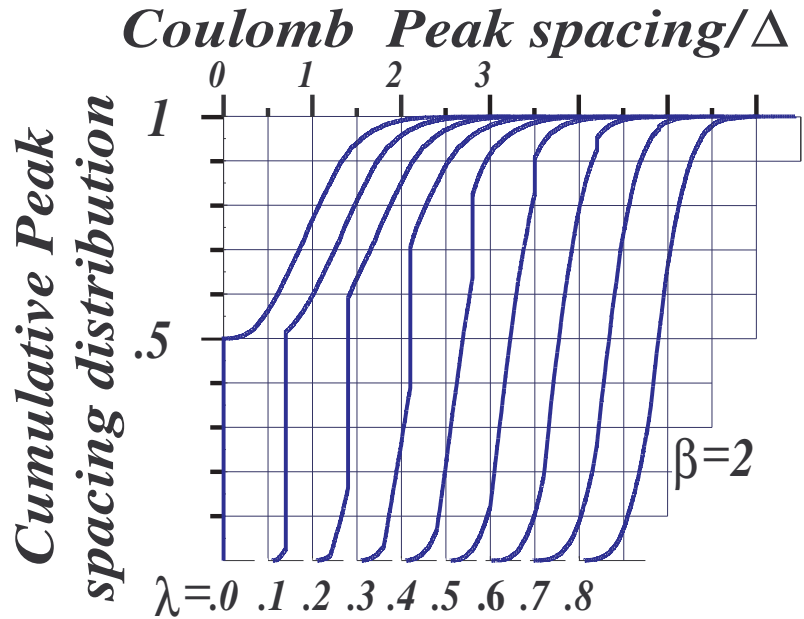}
    \vskip -1.5cm
\caption{Theory of cumulative peak spacing distribution in the
absence of time reversal symmetry ($\beta=2$). We assume that the
external magnetic field is large enough to break time reversal
symmetry, but small enough so that we can neglect Zeeman
splitting. In other words, the interaction in the Cooper channel is
completely suppressed and the level spacing distribution is
described by a GUE ensemble. For clarity we shifted the
distributions with exchange parameters $\lambda=-J_{\rm s}/\Delta$
by intervals of half level spacing.\label{fg:bne0the}}
\end{figure}

 Sequence $\# 3$ will occur if $E_{2,0} < E_{2,1}$ and $E_{4,0}<
E_{4,1}$. This occurs if $J_{\rm c}-2J_{\rm s} < \delta < \tilde
\delta 2J_{\rm s} - 3J_{c}$ and leads to a step function jump in
the cumulative peak distribution at $\Delta E_3$. This is clearly
seen in the theoretical curves of Fig.~\ref{fg:b0the} and
Fig.~\ref{fg:bne0the}. The weight of the jump is bounded from
above by $\int_{\Delta E_3}^\infty p(\delta')d \delta'$ where
$p(\delta')$ is the Wigner distribution for consecutive levels at
distance $\delta'$.\\ Sequence $\# 5$ requires that $E_{2,1} <
E_{2,0}\; {\rm and }\; E_{4,0} <  E_{4,1}$, which occurs when
$\delta < \min\{J_{\rm c}-2 J_{\rm s},\tilde \delta -3J_{\rm c}+2
J_{\rm s}\}$. This leads to a low energy tail in the shape of the
Wigner distribution curve, that extend all the way down to $\Delta
E =0$, when $J_{\rm c}=0$.

Here we have discussed some of the simplest spin-entry sequences,
including situations with total spin $0$, $1/2$, and $1$, and we
have seen how these sequences appear in the cumulative peak
spacing distribution. Generalizations for more complex situations
are straightforward but tedious. Different sequences will lead to
other singularities and smooth curve-segments in the distribution
of peak spacings.

The actual numerical calculations (which includes also sequences
with spin larger than $1$) do not demand, however, a detailed
analysis as above. We performed them in the following manner: we
use 24 random levels around the center of the spectrum of a random
$100 \times 100$ matrix for 1000 realizations. For each
realization we find, using formula (\ref{eq:EGdiff}), the
configuration of the groundstate and its energy for eight
consecutive entries of electrons, this gives us 6 peak spacings
for each realization, so that totally we plot a histogram of 6000
peaks.

In Fig.~\ref{fg:bne0the} we plot the cumulative peak spacing
distribution for magnetic field large enough so that the non
interacting levels may be described by the GUE ensemble and the
interaction in the Cooper channel is completely suppressed. (We
assume that the magnetic field is small enough so we can neglect
Zeeman splitting effects.)

\subsubsection{ Comparison between Theory and Experiments}

\label{se:exp} This section compares our theory for Coulomb
blockade peak spacing with the available experimental results. We
will see below that the agreement between theory and experiment is
not very good, particularly for small peak spacings. However some
features of the theory are found also in experiments. For example,
a non-Gaussian tail at large values of peak spacing and a jump in
the cumulative distribution, that we described in the preceding
section, are present in few experiments.

\begin{figure}[ht]
    \vglue 0cm \epsfxsize=0.8\hsize
    \hspace{0.1\hsize}
    \epsffile{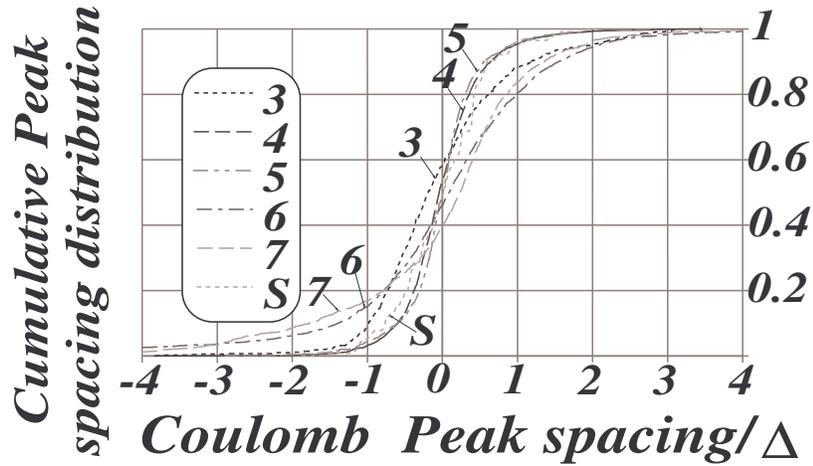}\vspace{-0.7cm}
\caption{ \label{fg:b0CharEns}
  Experimental results (curves  ``3-7'' \protect{\cite{WF:Patel98}}
and ``S'' \protect{\cite{WF:Luscher01}}) of the cumulative peak
spacing distributions. To compare between the dots of different
sizes, we subtract from each distribution the dot average peak
spacing and scaled it to the average level spacing, $\Delta$. The
dots here have time reversal symmetry.}
\end{figure}

Fig.~\ref{fg:b0CharEns} depicts the cumulative peak spacing
distribution of AlGaAs-dots. Dots ``3-7'' were studied in
Ref.~\cite{WF:Patel98} and dot ``S" in \cite{WF:Luscher01}. In all
the experimental curves that we present here, no magnetic field is
applied. We therefore assume in the theoretical analysis that time
reversal symmetry is conserved. In each curve we normalize the
peak spacing by the dot level spacing, which vary from dot to dot
since their sizes are different. We have used here the average
level spacings quoted in the experimental papers, there are, of
course, some uncertainties in these values. After this
normalization we would expect that the curves of dots 3-7 will be
similar. This should happen because they have similar electron
density and therefore similar $r_{\rm s}$ and exchange interaction
parameter (see Appendix~\ref{ap:LP}). In addition we would expect
that dot "S" will have a different curve, because it has an
electron density that is larger by a factor $\sim 3$. However, as
Fig~\ref{fg:b0CharEns} shows, the experimental results behave
differently. This may be attributed to the intrinsic noise in the
system (see Table~1 in \cite{WF:Patel98}), but we still do not
understand completely the origin for this behavior.
\begin{figure}[ht]
    \vglue 0cm \epsfxsize=0.8\hsize
    \hspace{0.1\hsize}
    \epsffile{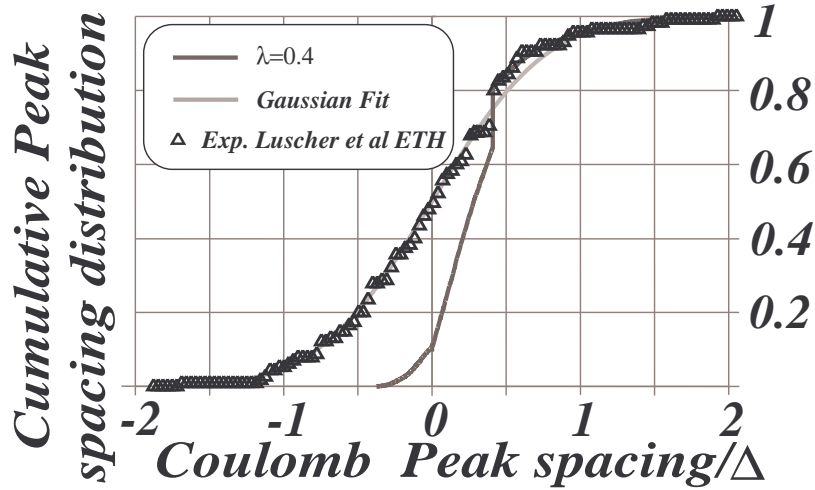} \vspace{-0.7cm}
\caption{ \label{fg:etb0}
  Comparison between theory and experiment\protect\cite{WF:Luscher01}.  To fit best the upper tail of
  the experimental results in the absence of magnetic field, we
  choose in the theory (with GOE symmetry but without interaction
  in the Cooper channel) $-J_{\rm s}/\Delta =\lambda=0.4$.}
\end{figure}
The data of Refs.~\cite{WF:Sivan96} and \cite{WF:Simmel97} show a
significantly wider distribution of level spacings than found in
Ref.~\cite{WF:Patel98}, despite the apparent similarity of the
systems studied by these  groups. So far, there has been no clear
explanation for this discrepancy.

 We nevertheless plot in Fig.~\ref{fg:etb0} our theory and the
experimental curve of \cite{WF:Luscher01} (without magnetic
field). The overall experimental curve fits well to a Gaussian.
However, the details of the upper (right) tail of the cumulative
distribution, {\em i.e.}, the jump and the non Gaussian tails fits
better to the theory of the spin fluctuations in the Ground state.
To fit best the upper tail of the experimental results in the
absence of magnetic field, we choose in the theory (with GOE
symmetry but without interaction in the Cooper channel) $-J_{\rm
s}/\Delta =\lambda=0.4$. Notice that using the static RPA estimate
of $\lambda$ (in Appendix~\ref{ap:LP}), with the experimental
value $r_{\rm s} \sim 0.72$ ~\cite{WF:Luscher01} we find, as
expected, a value that is somewhat smaller than $0.4$.

We expect that several effects, (not included in the theory) may
be important for current experiments. The lower part of the peak
spacing distribution, that is built from single particle levels
that are, by chance, very close to each other, is especially
sensitive to these effects. Indeed, this part appears to be far
from the theoretical curves. Among them are:
 \begin{enumerate}
  \item Non universal effects of finite $g$ that cause fluctuations
 in the interaction parameter \cite{WF:Ullmo01}. For ballistic
two-dimensional dots $g=\sqrt{2 \pi n A}$, where $A$ is the dot
area and $n$ is the density of the electrons. In the dots of
Ref.~\cite{WF:Sivan96,WF:Simmel97,WF:Patel98} $r_{\rm s} \sim 1-3$
and $g \sim 50 -150$ (in these of Ref.~\cite{WF:Luscher01} $r_{\rm
s} \sim 0.72$ and $g \sim 35$). Thus, when we compare theory to
experiment, fluctuations in the interaction constants cannot
always be ignored. Ullmo and Baranger present in
Ref.~\cite{WF:Ullmo01} a detailed study of the effect of
fluctuations of the interaction parameters (see also the
discussion here in page \pageref{pg:flactuations}). They find
indeed that when these fluctuations are included the results ``are
significantly more like the experimental results than the simple
constant interaction model''.
 \item We assume that the temperature, and the single
particle levels width (due to tunneling to the leads), is smaller
than the mean level spacing and therefore we ignore their effects.
This assumption is not valid for the lower part of the
distribution as it is built from levels whose distance from
neighboring levels might be much smaller than the average level
spacing.\\ The importance of the temperature was considered very
recently by Usaj and Baranger \cite{WF:Usay01}. They find that
temperature effects are significant even at $T \sim 0.1 \Delta$.
\item There is experimental noise due to charge motions during the measurements time.
This effect \cite{WF:Patel98} might be the dominant contribution
to the smearing of the distribution in the experimental curve.
\end{enumerate}

\section{Spin-Orbit Effects}
\label{se:SO}

Spin-orbit coupling can have major effects on the ground states or
the low-energy transport properties of a mesoscopic system.  In
many  metallic nanoparticles, spin-orbit  effects arise from
randomly placed heavy-ion impurities, which can simultaneously
scatter electrons and flip their spin, subject to the constraints
imposed by the requirement of time-reversal invariance in the
absence of an applied magnetic field.  In other cases, one is
concerned with metal particles where the spin-orbit effects are
already significant in the band-structure of the ideal host
crystal, so that the ``spin'' variable in the Bloch states
actually represents a mixture of spin and orbital degrees of
freedom at the microscopic level.  In this case spin-flip
scattering with the requisite spin-orbit symmetry can occur
whenever there is scattering: from defects, from impurities, or
from the boundaries of the sample.  Spin-orbit scattering in the
above cases can generally be characterized by a spin-orbit
scattering rate, and the importance of spin-orbit effects is
determined by the ratio of this rate to other frequencies
characteristic of the mesoscopic system. Effects of spin-orbit
scattering on the groundstate spin-structure and on the energy
splitting in an applied magnetic field will be discussed in
Subsection~\ref{se:gtens}.  The effects of spin-orbit coupling on
the spacing of groundstate energies, in the presence of
electron-electron interactions, will be discussed in
Subsection~\ref{se:so+int}.

A peculiar situation can arise in two-dimensional electron systems
in materials such as GaAs.  Here the dominant spin-orbit  effects
arise from terms in the effective Hamiltonian in which there is a
coupling to the electron spin linear in the electron velocity.
(These terms arise from the asymmetry of the potential well
confining the electrons to two dimensions and from the lack of
inversion symmetry in the GaAs crystal structure.)  The special
form of this coupling leads to a large suppression of spin-orbit
effects when the 2D electron system is confined in a small quantum
dot.  However, effects of spin-orbit coupling are again  enhanced
in the presence of a strong magnetic field parallel to the plane
of the sample, so that they must be taken into account in such
properties as the level-spacings of a closed dot or the statistics
of conductance fluctuations in a dot coupled to leads through one
or more open channels.  These effects will be discussed in
Subsection~\ref{se:GaAs} below.
%%
%%
%%%%%%%%%%%%%%%%%%%%%%

\subsection{Effective $g$-tensor of a metal particle with spin-orbit scattering.}
\label{se:gtens}

According to Kramers theorem, a metal particle with an odd number
of electrons with no special symmetry, in zero magnetic field,
must have a degenerate groundstate manifold, with pairs of states
related to each other by time-reversal symmetry.  In the absence
of spin-orbit coupling, the total spin $S$ is a good quantum
number, and the groundstate manifold is just that expected for
half-integer $S$.  As we have seen in Sec.~\ref{se:interactions},
if the electron-electron interaction is weak, we will essentially
always find $S=1/2$ for odd $N$ and the ground state will be just
two-fold degenerate.  For stronger electron-electron interactions,
however, there will be some probability of finding $S=3/2$ or
larger, so that four-fold or higher degeneracies are also
possible.  When spin-orbit interactions are turned on, the higher
degeneracies will be broken into a set of doublets, so that the
ground state will again be two-fold degenerate.

If we now apply a magnetic field  $B$ to the system, the
degenerate ground state will be split.  For sufficiently small
$B$, one of the states will move up in energy by an amount $\delta
\varepsilon$ which is linear in $B$, while the other state will
move down by the same amount.  These shifts may be measured, at
least in principle, by electron-tunneling spectroscopy experiments
in an applied magnetic field.   We discuss here the statistical
properties of the distribution of energy shifts expected under
various circumstances. We concentrate on the situation where the
electron-electron interaction is weak, so that the many-body
ground state is well described by the picture of weakly
interacting quasiparticles, as effects of electron-electron
interactions will be discussed in the next subsection.

Quite generally, we may write the linear splitting of a Kramers doublet in
the form
\begin{equation}
\label{eq:kramerdoublet} \delta \varepsilon = |\mu_B /2| (\vec{B}
\cdot \tensor{K} \cdot \vec{B})^{1/2}
\end{equation}
where $\mu_B<0$ is the electron Bohr magneton and $\tensor{K}$ is
a real, positive-definite  symmetric $3\times 3$ tensor.  In the
absence of spin-orbit coupling, $\tensor{K}$ is isotropic, with
$K_{ij}=4 \delta_{ij} $.  When spin-orbit coupling is present, we
find that $\tensor{K}$ varies from level to level, and is in
general anisotropic.  We write the three eigenvalues of
$\tensor{K}$ as $g_k^2$, ($k=1,2,3$), with $|g_1| \le |g_2| \le
|g_3| $, and refer to the $g_k$'s as the three principal
$g$-factors for the level. Although the energy-splittings in a
static magnetic field only define the absolute values of the
$g_k$, by considering the response to a time-varying magnetic
field (e.g., a spin resonance experiment) one can also give an
unambiguous meaning to the sign of the product of the three
$g$-factors.  Since the sign of an individual $g_k$ has no
physical meaning, we adopt the convention that  $g_3$ and $g_2$
are always positive, but $g_1$ can be positive or negative,
depending on the specific system considered.

For the case of weakly interacting electrons, which we consider
here, the ground  state for $2N+1$ electrons consists of  $2N$
electrons in filled Kramers doublets, plus one electron in a
doublet which is singly occupied. The filled doublets give no
contribution to the linear energy shift because in each case one
state moves up and the other moves down by the same amount.  Thus,
the $g$-factors are determined by the properties of the
singly-occupied state.

In the presence of spin-orbit coupling there are  two
contributions which can shift the $g$-values from the bare value
$g=2$.  If we take into account only the interaction of  $B$ with
the electron spin, then spin-orbit coupling will always reduce the
$g$-values.  For example, if the magnetic field is applied in the
$z$-direction, the state which is shifted down in energy will be
the particular linear combination of the two degenerate states
which has the maximum expectation value of $-S_z$.  This
expectation value is $\le 1/2$, so the spin-contribution to the
$g$-factor will generically be reduced by spin-orbit coupling.

On the other hand, there is also an orbital contribution to the
linear Zeeman effect, when spin-orbit coupling is present . (In
the absence of spin-orbit coupling, the orbital states in an
irregular dot will be generically non-degenerate and
time-reversal-invariant, so they cannot acquire a linear energy
shift in a weak magnetic field.) Both the orbital and spin
contributions were considered by Matveev {\it et al.}
\cite{SOC:Matveev00}, who discuss the expectation value and
probability distribution of $\delta \varepsilon^2$ for the
magnetic field in an arbitrary fixed direction.

\begin{figure}[ht]
\hspace{0.05\hsize} \epsfxsize=0.9\hsize \epsffile{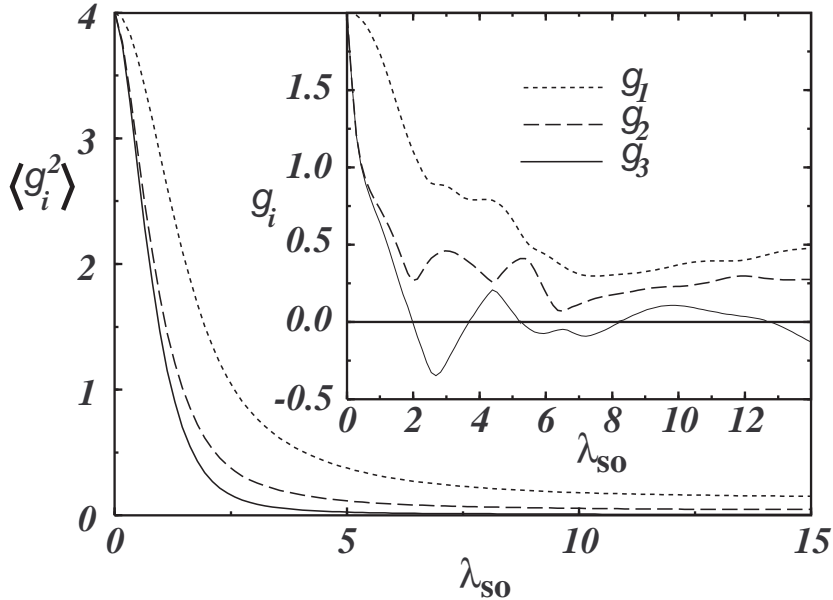}
\caption{\label{fg:gtens} Average of the squares of principal
$g$-factors versus spin-orbit scattering strength $\lambda_{\rm
so}$, obtained from numerical simulation of a random matrix model.
Inset: $g_1$, $g_2$, and $g_3$ for a specific realization. We have
included the sign of $g_1$.}
\end{figure}

By contrast Brouwer {\it et al.}~\cite{SOC:Brouwer00} considered
the joint probability distribution of the three $g$-values for a
single level, so they could examine the anisotropy as well as the
magnitude of the $g$-tensor.  Their analysis concentrated on the
case where orbital effects can be ignored, so that the mean-square
$g$-factors are monotonically reduced with increasing spin-orbit
coupling.  The strength of the spin-orbit coupling in this case is
determined by a parameter
\begin{equation}
\lambda_{\rm so} = \sqrt{\frac{\pi \hbar}{\tau_{\rm so} \Delta}}
\end{equation}
where $\Delta$ is the mean separation between one-electron energy
levels.   The mean spin-orbit scattering time $\tau_{\rm so}$ is
defined so that if we prepare a state with spin up, the
probability to find it in the same spin direction after time $t$
is $\sim e^{-t/\tau_{\rm so}}$. When $\lambda_{\rm so} \gg 1$, one
finds that the $g$-factors are greatly reduced from their bare
values, and one can obtain an analytic form for the joint
probability distribution:
\begin{equation}
\label{eq:jointprob} P\left(g_1,g_2,g_3 \right) \propto
\prod_{i<j}\left|g_i^2-g_j^2\right|\prod_{i}e^{-3g_i^2/2\left\langle
g^2\right\rangle}
\end{equation}
For intermediate values of the coupling parameter $\lambda_{\rm
so}$, one can perform numerical simulations to study the
distribution, using random-matrix theory.  In Fig.~\ref{fg:gtens}
we show the $\lambda_{\rm so}$-dependence of the mean values of
$g_k^2$, as well as the values of $g_k$ for a particular
realization of the random matrices.

Very recently, Petta and Ralph [Ref.~\cite{SOC:Petta01}] have
measured effective $g$-values for a number of levels in each of
several different nanoparticles, of Cu, Ag and Au, with diameters
in the range $3 \mbox{ - }5nm$.  They did not vary the direction
of the applied magnetic field, so they could not study the
anisotropy of the $g$-tensor. However, the statistical
distributions of the $g$-factors (normalized to their means) for
different levels in a given particle were found to be in good
agreement with the theories of Refs.~\cite{SOC:Matveev00} and
\cite{SOC:Brouwer00}. For the mean $g$-factor, an agreement with
Refs.~\cite{SOC:Matveev00} and \cite{SOC:Brouwer00} is found if
the spin contribution is taken into account only; the mean
$g$-values observed in the Au particles (ranging from 0.12 to
0.45) were significantly smaller than what one might expect from
the orbital contribution, according to the theory of
Ref.~\cite{SOC:Matveev00}, unless one assumes a very short
mean-free-path for the electrons. (Using the formulas in
Ref.\cite{SOC:Matveev00}, one would need a mean-free-path of order
$0.1 nm$ to get $g$-values this small.) Small $g$-values (below
0.5) for Au nanoparticles were also observed previously by
Davidovic and Tinkham~\cite{WF:Davidovic00}.
%
%
%%%%%%%%%%

\subsection{Effects of spin-orbit coupling on interaction-corrections
to groundstate configurations and energy spacings.}
\label{se:so+int}

So far, we have ignored the effects of electron-electron
interactions. This should generally be valid if the exchange
interaction is small compared to the threshold for the Stoner
instability, so that the probability of finding $S>1/2$ is small
in the absence of spin-orbit coupling.   When the spin-orbit
coupling parameter $\lambda_{\rm so}$ is large, the effective
exchange interaction between two electrons in states close to the
Fermi-energy of the particle will be even further reduced, as the
mean spin in any state becomes small compared to $1/2$, and the
local spin orientations have different spatial distributions for
one-electron states belonging to different Kramers doublets.

In the limit of very large spin-orbit coupling, where the mean
spin tends to zero, the exchange interaction should also tend to
zero. This means that the parameter $J_{\rm s}$ in the effective
Hamiltonian (\ref{eq:heff}) of Sec.~\ref{se:Heff} should be set to
zero. This is consistent with the fact that spin is no longer a
good quantum number of the system, and the term proportional to
$J_{\rm s}$ is no longer invariant under the set of allowed
unitary transformations of the random matrices.

A consequence of this analysis is that if a spin-orbit scatterers
are added to a system with fixed electron-electron interaction
(fixed $r_{\rm s}$) the probability distribution for the
separation of successive groundstate energies, measured by the
Coulomb-blockade peak separations, should approach that of a
non-interacting electron system in the symplectic ensemble.  This
means that there should be a bimodal distribution with an even-odd
alternation. The chemical potential to add a second  electron to a
Kramers doublet is the same as the energy to add the first
electron, after the coulomb blockade energy [$U_{\rm c}$ in
Eq.~(\ref{eq:heff})] is subtracted, whereas the chemical potential
for the next electron will be larger by an amount approximately
given by the mean level spacing~$\Delta$.

\subsection{Spin-orbit effects in a GaAs quantum dot in a parallel magnetic field.}
\label{se:GaAs}

The most important spin-orbit terms in the effective Hamiltonian for a 2D
electron gas (2DEG) in a GaAs heterostructure or quantum well may be
written in the form
\begin{equation}
\label{eq:hso} \mathcal{H}_{\rm so} = \gamma_1 v_x \sigma_y -
\gamma_2 v_y \sigma_x
\end{equation}
where $\vec v$ is the electron velocity operator. We have assumed
that the 2DEG is grown on a [001] GaAs plane, and we  have chosen
the $x$ and $y$ axes to lie in the [$110$] and [$1\bar{1}0$]
directions. For an open 2DEG  this leads to a spin-orbit
scattering rate of order $\gamma^2 D$, where $D$ is the diffusion
constant and $\gamma$ is the geometric mean of the two coupling
constants in Eq.~(\ref{eq:hso}).  For a confined dot of radius
$R$, in zero magnetic field, however, the effects of spin-orbit
coupling are suppressed  if the typical angle of spin precession
for an electron crossing the dot, given by $\theta = \gamma R$, is
small compared to unity.  One finds in this case that the matrix
elements of $\mathcal{H}_{\rm so}$ are greatly reduced for energy
states whose energy separation is small.

Halperin {\it et al.} \cite{SOC:Halperin01} have argued that the
effects of spin-orbit coupling can be enhanced, however, in the
presence of an applied magnetic field in the plane. The
enhancement is maximum when the Zeeman energy becomes comparable
to the Thouless energy ({\em i.e.}, the inverse of the transit
time for an electron in the dot), in which case there is an
effective spin-mixing rate comparable to the spin-orbit scattering
rate for an open system with an electron mean free-path equal to
the mean free path in the dot.  For a closed dot, the spin-mixing
would be manifest in the repulsion of energy levels for different
spin, and the appearance of anti-crossings of the levels as when
the Zeeman field is varied.

Motivated by experiments of Folk {\it et al.}~\cite{SOC:Folk01},
Halperin {\it et al.}~\cite{SOC:Halperin01} considered the
``universal conductance fluctuations'' of a dot connected to a
pair of leads with one or more channels open in each lead.  They
considered explicitly the case where there is a weak magnetic
field perpendicular to the dot, so that time reversal symmetry is
broken, and the system is in  the class of the unitary ensemble,
even in the absence of spin-orbit coupling. It was shown that
effects of spin-orbit coupling in large Zeeman field could then
lead to a factor of two reduction in the variance in the
conductance, which is in addition to the factor of two reduction
caused by breaking of the spin degeneracy.  Calculations of the
cross-over, as a function of the in-plane magnetic field, were in
at least qualitative agreement with the experimental observations.

Very recently, Aleiner and Fal'ko\cite{SOC:Aleiner01} have
considered the case without a perpendicular magnetic field, so
that the system without spin-orbit coupling would be in the
orthogonal ensemble.  They have shown that the application of a
parallel magnetic field in this case turns on a spin-orbit
perturbation with a special symmetry, so that the system retains
an effective time-reversal symmetry even in the presence of the
large Zeeman field. The spin-orbit coupling leads to a reduction
in the size of conductance fluctuations, but not as much as one
would obtain if the time-reversal symmetry was also broken.  The
spin-orbit coupling also leads to a reduction in the ``weak
localization'' correction to the average conductance,  but does
not lead to complete suppression as one would find for a broken
time reversal symmetry. (However, as noted by Meyer {\it et al.}
\cite{SOC:Meyer01} and  by Fal'ko and Jungwirth\cite{SOC:Falko01},
for an asymmetric quantum well of finite thickness, application of
a strong magnetic  field parallel to the sample can lead to broken
time-reversal symmetry due to orbital effects, even in the absence
of spin-orbit coupling.)

%%
%%
%%%%%%%%%%%%%%%%%%%%%%%%%%%%%%%%%%%%%%%%%%%%%%%%%%%
%%  MULTIPLETS
%%
%%
%%
%%
\section{Origin of multiplets in the differential
conductance} \label{se:mutiplets}

In a recent experiment  Davidovic and Tinkham\cite{WF:Davidovic00}
studied tunneling into individual Au nanoparticles of estimated
diameters $2$–$5nm$, at dilution refrigerator temperatures. The
differential conductance $dI/dV$, as a function of the
source-drain voltage $V$, indicate resonant tunneling via discrete
energy levels of the particle. Unlike previously studied normal
metal particles of Au and Al, in these samples they find that the
{\em lowest} energy tunneling resonances are split into clusters
of 2-10 sub-resonances. The distance between resonances within one
cluster is much smaller than the mean level spacing of the Au
grain.

\begin{figure}[ht]
\hspace{0.15\hsize}
\epsfxsize=0.7\hsize
\epsffile{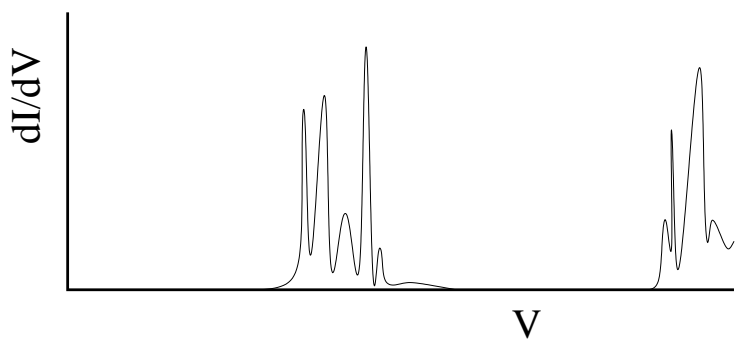}
\caption{\label{fig:illustr} Schematic illustration of the
experimental results of Ref.~\protect\cite{WF:Davidovic00}: The
peaks in the differential conductance are split. The distance
between the multiplets is of the order of the single particle
level spacing $\Delta$; the distance between peaks in the same
multiplet is much smaller.}
\end{figure}

This situation is illustrated schematically in
Fig.~\ref{fig:illustr}. The differential conductance $dI/dV$ shows
resonances, where each resonance in $dI/dV$ is actually a
multiplet, the splitting between the peaks of the multiplets being
a factor $\sim 30$ smaller than distance between the resonances
(which is of the order single-particle level spacing in the
grain).

In this section we outline two-different mechanisms which can lead
to a fine structure of the first conductance peak. We first show
how such a fine structure can occur if the ground state has a
finite spin with small energy splittings between states of
different magnetic quantum number. In this model it is necessary
to have a relatively large total spin in order to split the
conductance peak into many sub-peaks. This mechanism would also be
suppressed by large spin-orbit coupling.

In the second mechanism, following Agam {\em et al.}
\cite{WF:Agam97}, we show how such a fine structure can arise from
nonequilibrium processes induced by the large bias voltage $V$
used in the experiment. This mechanism seems to us to be the more
likely one for explaining the observations of
 Ref.~\cite{WF:Davidovic00}.  We also indicate how,
experimentally, one might distinguish between the two proposed
explanations.

\subsection{Multiplets from an almost degenerate groundstate}
\label{se:almost_degenerate}

In general, a peak in the differential conductance as a function
of the bias voltage $V$ may occur if an additional channel for
tunneling onto or from the metal grain is opened at that $V$. The
relation between $V$ at the peaks and the ground state energies is
complicated; it depends on the capacitive division between the
left and right contact and on the conductances of the two
tunneling contacts. A detailed account of the possible scenarios
can be found in the review by von Delft and
Ralph~\cite{vonDelftRalph}.

Here, we make the simplifying assumption that the left point
contact has the bigger resistance and the smaller capacitance, so
that the electrostatic potential of the dot equals that of the
right reservoir, and the contact to the left reservoir can be seen
as the ``bottleneck'' for current flow. Then, if the grain has $N$
electrons at zero bias, a conductance peak occurs when $$ eV =
E_{N+1} - E_{N}, $$ {\em i.e.}, when the bias voltage $V$ is
precisely equal to the difference of the energies of any two
many-body states of the grain with $N$ and $N+1$ electrons (for $V
> 0$), provided the initial $N$-particle state is populated at the
corresponding bias. Below we focus on the first peak in the
differential conductance and discuss when and how a fine structure
of that first peak can arise.  We assume that the temperature is
small compared to any splittings in the energy levels, and we
assume (for the moment) that there is no spin-orbit coupling.

\begin{figure}[ht]
\epsfxsize=\hsize \epsffile{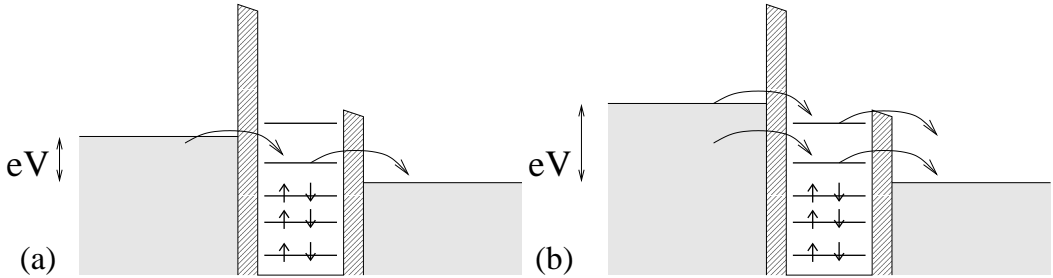}
\caption{\label{fig:reservoirs} Schematic drawing of the tunneling
process considered here. The left point contact has the smaller
capacitance and the smaller conductance. When the bias voltage is
increased, peaks in the differential conductance occur, whenever a
new channel for tunneling onto or from the grain is opened.
Compare the bias voltages in (a) and (b).}
\end{figure}

\begin{figure}[ht]
%\hspace{0.15\hsize}
\epsfxsize=\hsize \epsffile{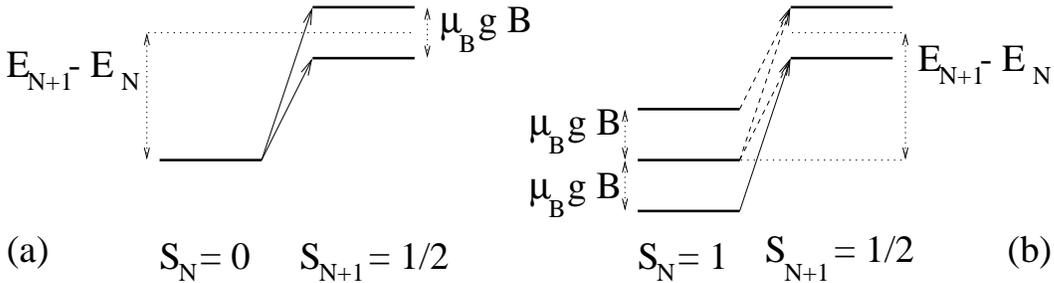} \caption{\label{fig:1}
Possible transitions between Zeeman split states with $N$ and
$N+1$ particles, for $S_N=0$, $S_{N+1}=1/2$ (a), and $S_{N}=1$ and
$S_{N+1}=1/2$ (b). Note that the transitions starting out of the
excited states of the triplet in (b), denoted by the dashed
arrows, do not give rise to peaks in the differential conductance,
(assuming that equilibrium is reached between successive tunneling
events), because the excited states are not populated at $eV =
E_{N+1} - E_{N} - \mu_B g B/2$. }
\end{figure}

If the $N$ and $N+1$-particle ground states have perfect
degeneracies, the difference $E_{N} - E_{N+1}$ can only take a
single value, and a single peak will be observed, no matter what
the spin of the ground state is. Hence to observe a fine
structure, the degeneracy of the ground state has to be lifted.
This can be done by application of a uniform magnetic field, as is
illustrated in Fig.~\ref{fig:1} for $S_N = 0$ or $S_N = 1$ and
$S_{N+1} = 1/2$. In the case where $S_{N+1} = S_N + 1/2$, the
difference $E_{N+1} - E_{N}$ between the energies of the many-body
states for $N$ and $N+1$ particles can take two values,
$$ E_{N+1} - E_{N} = E_{N+1}^{0} -
E_{N}^{0} \pm (1/2) g \mu_B B, $$ where $E_{N}^{0}$ and
$E_{N+1}^{0}$ are the $N$ and $(N+1)$-particle energies in the
absence of the magnetic field.  The differential conductance shows
a double peak at voltages $$ e V_{\pm} = E_{N+1}^{0} - E_{N}^{0}
\pm (1/2) g \mu_B B, $$ as is seen in Fig.~\ref{fig:1}a. On the
other hand, only a single peak at bias voltage $V_{+} = E_{N+1} -
E_{N} + (1/2) g \mu_B B $ is found if $S_{N+1}= S_{N} - 1/2$.
Although the bias voltage $V_{-}$ corresponds as well to a
transition energy between many-body states with $N$ and $N+1$
particles for $S_{N} > S_{N+1}$, no peak in the differential
conductance is found at that bias voltage, because the initial
state of that transition is an excited state, which is not
populated at $V = V_{-}$. Population of an excited $N$-particle
state is only possible at higher bias voltages $V \ge V_{+}$ via
inelastic processes that use the $N+1$-particle state as an
intermediate step. (A small nonequilibrium population of the
excited $N$-particle state, and hence a small peak at $V = V_{-}$,
may, however, occur as a result of inelastic cotunneling, as is
explained in Ref.~\cite{AleinerAgam}.) If the difference in the
total spin quantum numbers for $N$ and $N+1$ is greater than 1/2,
then there can be no conduction peak at all in the absence of
spin-orbit coupling or inelastic cotunneling processes.

The situation changes in the presence of weak static magnetic
impurities. ``Weak'' here means that they can be seen as a small
perturbation on top of the picture sketched in the previous
sections.  ``Static'' means that the impurity ion has a large
intrinsic angular-momentum and large crystal-anisotropy, so that
we can neglect the matrix elements for transitions between
different impurity spin-states. The impurity spins could be in the
grain itself or could be located close to the grain in the
surrounding insulator. Then if the many-body state of the grain
has non-zero spin, the spin degeneracy will be lifted by the
coupling to the impurity spin, which can give rise to a splitting
of the lowest conductance peak even in the absence of an applied
magnetic field.  A significant difference between this case and
the splitting due to an external field is that the effective
coupling now depends on the microscopic details of the electron
wavefunctions close to the impurity, and the level splitting will
generally be different for the $N$ and $N+1$ electron states. As
we shall see, this makes it possible for the lowest conductance
resonance to split into more than two sub-peaks.

\begin{figure}[ht]
\hspace{0.15\hsize}
\epsfxsize=0.7\hsize
\epsffile{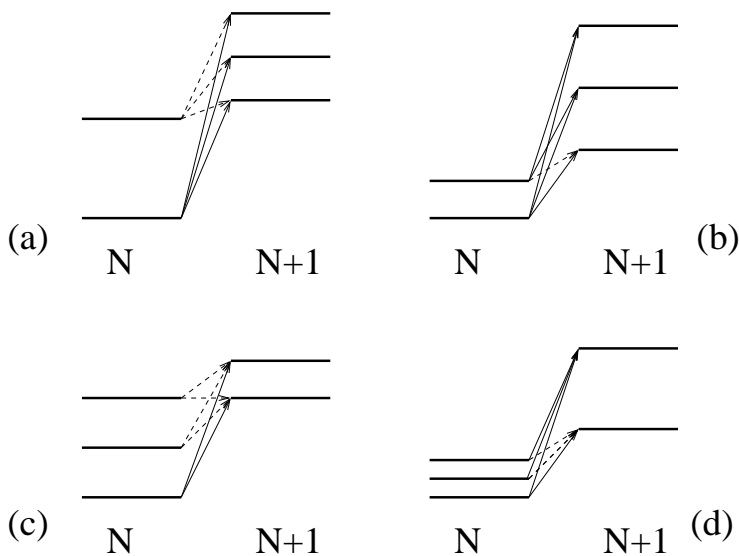}
\bigskip
\caption{\label{fig:2} Possible transitions between states with
$S_{N}=1/2$, $S_{N+1}=1$ (top) and $S_{N}=1$, $S_{N+1}=1/2$
(bottom), when the spin degeneracy is broken by the presence of
(several) static magnetic impurities. The transitions indicated by
solid arrows give rise to peaks in the differential conductance;
the transitions indicated by dashed arrows do not, because the
state they are starting from is not populated at the corresponding
bias voltage. How many peaks are visible depends on the actual
splitting of the energies. }
\end{figure}

We first consider the case of a single impurity spin.  According
to the Wigner-Eckart theorem, if the coupling to the impurity spin
is weak, an electronic many-body state with total spin $S$ will be
split into $(2S+1)$ equally-spaced levels, characterized by the
quantum number of the magnetic moment in the direction parallel to
that of the frozen impurity spin. The size of the splitting
depends on the concentration and microscopic details of the
impurities. Since a peak in $dI/dV$ can occur whenever $ eV =
E_{N+1} - E_{N}$, many close peaks appear when the degeneracy of
the ground state is lifted. The total number of possible
transitions is $(2 S_{N}+1)(2S_{N+1}+1)$, since now $E_{N}$ and
$E_{N+1}$ can take $2 S_{N}+1$ and $2 S_{N+1} + 1$ values,
respectively. However, for the same reasons as discussed above,
not all possible transitions give rise to peaks in the
differential conductance: Only transitions at energy differences
$\Delta E = E_{N+1} - E_{N}$ where the initial $N$-particle state
is already populated at a bias voltage $e V \le \Delta E$ are
reflected as peaks in the differential conductance, and the spin
component parallel to the frozen impurity spin can only change by
$\pm 1/2$. Some examples are shown in Fig.~\ref{fig:2} for $S_{N}
= 1/2$, $S_{N+1}=1$ and $S_{N}=1$, $S_{N+1}=1/2$. In the figure,
the transitions that correspond to true peaks in $dI/dV$ are shown
as solid arrows, the other ones are shown with dashed arrows. In
practice, since $e V$ is typically much bigger than the fine
structure of the $N$ and $N+1$-particle levels, all transitions
appearing at energy differences $\Delta E$ bigger than the
difference $e V_{\rm th} = E_{N+1}^{g} - E_{N}^{g}$ between the
ground state energies for $N$ and $N+1$ particles will show up as
true peaks at $e V = \Delta E$ , while no peaks appear for $e V <
e V_{\rm th}$ ($V_{\rm th}$ is the threshold voltage for current
flow).

If there are several frozen impurity spins coupling to the
electrons, we can again use the Wigner-Eckart theorem, in the case
of weak coupling, to show that a groundstate with spin $S$ is
split into $(2S+1)$ equally spaced levels classified by the spin
component along some direction which is a weighted vector sum of
the several frozen impurity spins. The weights in this sum will be
different in the states with $N$ particles and with $N+1$
particles, so that the quantization axes will generally be
different in the two states, as well as the level splittings. This
will lift the selection rule that the quantum number can only
change by $\pm 1/2$ on the addition of a single electron, so the
number of lines in the multiplet may increase accordingly.

The model of coupling to one or more frozen spins can also be
generalized to the case of dynamical spins. For example, in the
case of coupling to a single dynamical localized spin $S_{i}$ in
the insulating material close to the grain, the localized spin and
the spin of the electrons in the metal grain $S$ together will
form states with total angular momentum ranging from $|S - S_{i}|$
to $|S + S_{i}|$, and will split up in the corresponding
multiplet. In this case the number of possible transitions depends
on the detailed selection rules governing transitions of the
localized spin.

Finally, we consider the situation where spin-orbit coupling is
present.  In the case of weak spin-orbit coupling, a groundstate with
spin greater than $1/2$ can be split into several different energy
levels even in the absence of an applied magnetic field and in the
absence of magnetic impurities (although the splitting by spin-orbit
coupling only arises in second order perturbation theory, whereas the
splitting caused by magnetic impurities already appears in first order
perturbation theory). In general, the various
states will be split by different amounts, and so multiple subpeaks
can be observed, for large $S$, just as we found for the case with a
frozen magnetic impurity. In the present case, however, states with
odd $N$ remain twofold degenerate by Kramers' degeneracy, which
reduces the number of possible transitions roughly by a factor of two,
compared to the case of static magnetic impurities.

If the spin-orbit coupling is too strong, however, spin-orbit
splittings will become comparable to or larger than the
single-particle level spacings.  In this case, it is no longer
possible for splittings between spin states to give rise to fine
structure of the conductance peak on a scale small compared to
$\Delta$.  (Also, as we have seen previously in
Subsection~\ref{se:so+int}, exchange-splittings tend to be reduced
in this case, so that the one electron picture should be valid for
the ground states.) The gold particles studied in
Ref.~\cite{WF:Davidovic00} appear to be in the strong spin-orbit
coupling regime.

\subsection{Multiplets from nonequilibrium processes}
\label{se:nonequilibrium}

\begin{figure}[ht]
\hspace{0.15\hsize} \epsfxsize=0.7\hsize \epsffile{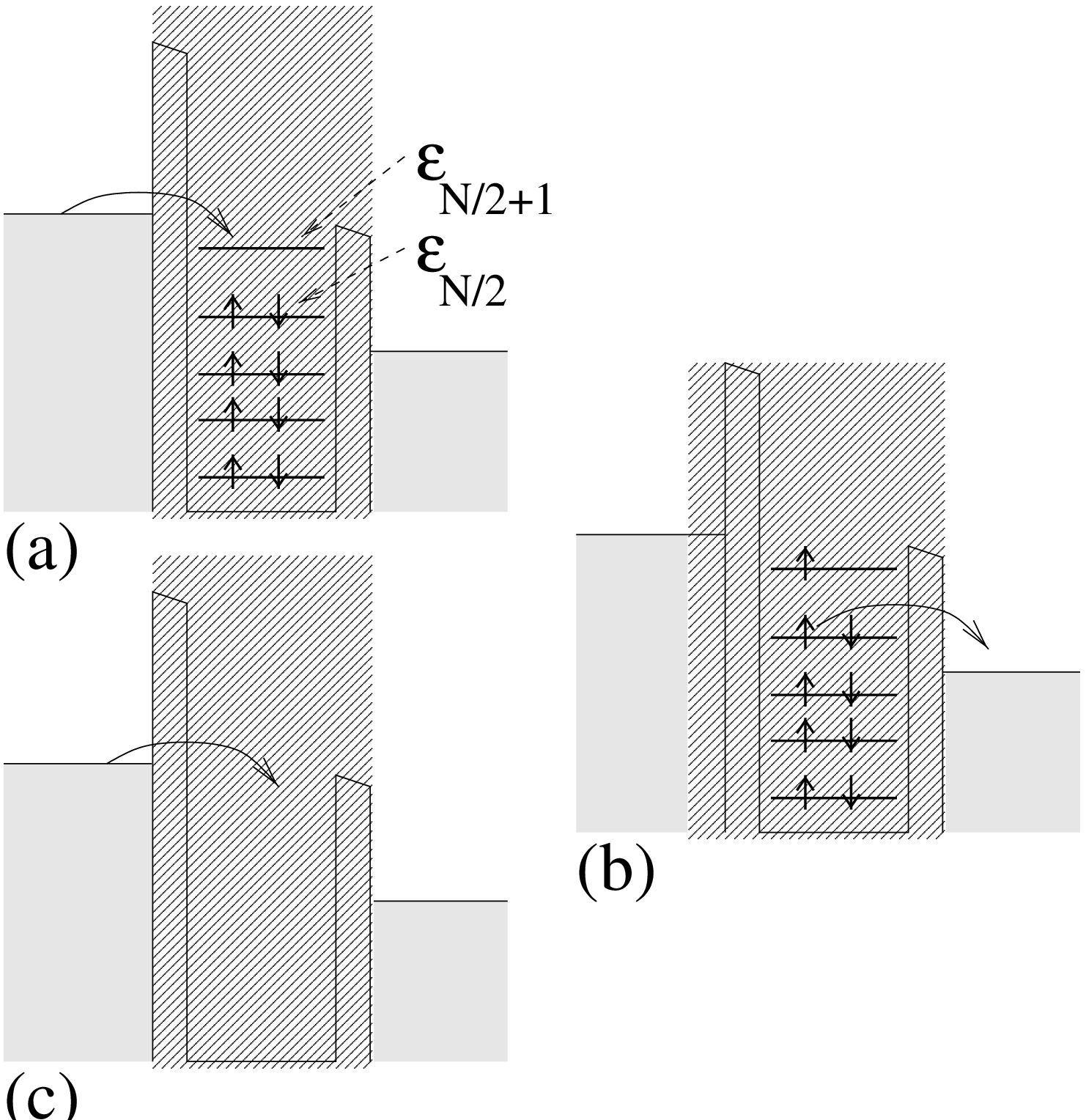}
\bigskip
\caption{\label{fig:noneq} A nonequilibrium configuration can be
obtained from the ground state (a) if an electron tunnels into the
first empty level (b), and another electron tunnels off the grain
from a different, lower lying, level (c). The energy required for
tunneling an electron into the highest level in the configurations
(a) and (c) may be different, which explains that more than one
peak can be seen in the differential conductance.}
\end{figure}

 A second mechanism to observe multiple
peak structures in the differential conductance is via
nonequilibrium population of highly excited states of the metal
grain, as was first proposed by Agam {\em et
al.}~\cite{WF:Agam97}. This mechanism does not need a degeneracy,
or near-degeneracy,  of the ground state. The idea of
Ref.~\cite{WF:Agam97} is as follows: Since the bias voltage is
typically much larger than the spacing $\Delta$ between single
particle levels, after an electron has tunnelled on and off the
grain, the grain may be left in an excited $N$-particle state,
with an occupation of the single-particle levels that differs from
the ground state, see Fig.~\ref{fig:noneq}. The fact that there is
a different occupation of the single-particle levels will slightly
shift the addition energies $E_{N+1} - E_{N}$, thus giving rise to
peaks in $dI/dV$ at different values of the bias voltage $V$. In
Ref.~\cite{WF:Agam97} spinless particles were considered. In that
case, nonequilibrium processes cause a fine structure of the
second and higher resonances, but not the first one
\cite{WF:Agam97}. For spin $1/2$ particles and even $N$, the
scenario of Ref.~\cite{WF:Agam97} can also lead to a fine
structure for the first resonance, as we will now describe.

We denote the highest occupied (self-consistent) single-particle
level in the $N$-particle ground state by $\varepsilon_{N/2}$ and
assume that the $N$-particle ground state has zero total spin.
When the bias voltage exceeds the threshold $e V_{\rm th} =
E_{N+1}^{g} - E_{N}^{g}$, current flow can leave the grain in an
excited $N$-electron state, when, after an electron has tunnelled
into the level $\varepsilon_{N/2+1}$, another electron tunnels out
of a lower-lying level $\varepsilon_{\nu}$, see
Fig.~\ref{fig:noneq}. Note that the excited state can have a total
spin $S=0$ or $S=1$. Since the grain is now in an $N$-particle
state that is different from the ground state (compare Figs.\
\ref{fig:noneq}a and c), the energy cost $E_{N+1} - E_{N}$ for
addition of an electron, and hence the position of a peak in
$dI/dV$, is, in general, different from $E_{N+1}^{g} - E_{N}^{g}$.
{\em A priori}, this difference can have three contributions: (1)
An electron can tunnel into different
%(and lower-lying)
single-particle levels than in the ground state. (2) The
transition energy $E_{N+1} - E_{N}$ depends on the spin $S$ of the
states involved, which can be different from the ground state
spin. (3) All transition energies depend uniquely (but weakly) on
the populations of the initial and final states through mesoscopic
fluctuations of the interaction contribution to the
energy~\cite{WF:Agam97}. The characteristic energy scales for the
first two of these contributions are $\Delta$ and $J_{\rm s}$,
while the third is of order $\Delta / \sqrt{g}$, as we shall see
below, where $g=2 \pi E_{\rm Th} /\Delta$ is the dimensionless
conductance of the grain.
For an even $N$, however, there exist  excited $N$-particle states
for which the level $\varepsilon_{N/2+1}$ is only singly occupied
and $S=0$, so that the first two contributions vanish. Then only
the contribution from mesoscopic fluctuations remains. If $g$ is
large, the energy scale for the fluctuations is small, and one
finds multiple conductance peaks close to $eV = E_{N+1}^{g} -
E_{N}^{g}$, where the total width of the multiplet is of order
$\Delta/\sqrt{g}$.

Although all nonequilibrium configurations have their own
characteristic transition energy $E_{N+1} - E_{N}$, not all of
them need to correspond to a peak in the conductance; no peak is
observed if the corresponding voltage is below the threshold
$V_{\rm th}$, which was needed to populate the corresponding
excited state. We estimate that the number of sub-peaks in the
first peak mutiplet, $N_{\rm first-peak}$ due to the nonequlibrium
effect is of order $eV/ 2\Delta$, which is roughly the ratio of
the Coulomb blockade energy to the single-particle level spacing.
%%%Note factor 2 in denominator.
%%%%%%%%More changes follow

To understand the reasons for this estimate let us first consider
the case where we can neglect the exchange coupling $J_s$ in the
effective Hamiltonian (\ref{eq:heff}).   (This is correct in the
case where spin-orbit coupling is strong). We also neglect the
Cooper-channel interaction $J_c$, for the reasons discussed in
Appendix C.  However, we take into account fluctuations in the
size of $U_c$ between different pairs of levels.  In this case,
the many-body states can be labelled by the occupancies of the
single-particle levels, and the ground states with energy
$E_{N}^{g}$ and energy $E_{N+1}^{g}$ are described by
Fig.~\ref{fig:noneq}a and b respectively. Let us assume that $V$
is close to,  but slightly above, the threshold voltage $V_{\rm
th}=E_{N+1}^{g} - E_{N}^{g}$. We further assume that when an
electron enters or leaves the system it does not excite other
electrons by multi-electron processes; in particular we ignore
Auger-like processes.  Similarly we assume that the only important
contributions to the conductance are via real, energy conserving,
transitions.

Under these assumptions, one finds that  the excited states that
contribute to $N_{\rm first-peak}$ have precisely one hole below
$\varepsilon_{N/2}$. The energy distance between these single-hole
states is roughly~$\Delta$. Since only excited states within an
energy $eV$ from the ground state can be populated, the number of
possible hole states is $\sim eV/\Delta$.  Of these, roughly half
will lead to positive energy shifts, which are necessary for
contributions to the multiplet structure, so we find $N_{\rm
first-peak} \sim eV/2\Delta$. The width of the peak is
proportional to the size of the fluctuations in $U_{\rm c}$ that
we have discussed in Subsection~\ref{se:fluctuations}; in case of
long range Coulomb interaction it is $\sim \Delta/\sqrt{g}$.

Many-body states with two or more holes below $\varepsilon_{N/2}$
cannot contribute to the conductance for voltages close to
$V_{\rm th}$, because if two holes are present when there are $N$
electrons on the particle, the level $\varepsilon_{N/2+1}$ will
necessarily be doubly occupied.  Then, the next electron would
have to enter through the level $\varepsilon_{N/2+2}$, which would
require an additional energy, of order $\Delta$, that is not
available for $V$ slightly above $V_{\rm th}$. If, by chance, the
distance of level $\varepsilon_{N/2+2}$ from level
$\varepsilon_{N/2+1}$ is smaller than $J_{\rm s}$ then transition
through nonequilibrium states with two holes may occur. This
situation will increase $N_{\rm first-peak}$ significantly, it
should be proportional now to $\sim (U/ \Delta)^2/2$.

Let us now consider the case where the exchange parameter $J_s$ is
not negligible.  Then, the $N$ electron states having one electron
in the level $\varepsilon_{N/2+1}$ and one hole below the Fermi
energy will have different energies depending on whether they have
$S=0$ or $S=1$.  For $S=1$, the energy is reduced by $|J_s|$, so
that the energy to add the next electron to the level
$\varepsilon_{N/2+1}$ is increased by $|J_s|$, and the peaks
arising from the triplet configuration will be shifted up by this
amount relative to the singlet contributions.  If this shift is
comparable to the level spacing $\Delta$, then only the singlet
peaks will appear close to the threshold, and the number of peaks
in the lowest multiplet will be the same as before, $N_{\rm
first-peak} \sim eV/2\Delta$. If $|J_s|$ is sufficiently small,
but not negligible, however, the singlet and triplet peaks may
appear to form a single multiplet, with twice as many peaks as
before.

\subsection{A comparison between the mechanisms of Subsections
\protect{\lowercase{\ref{se:almost_degenerate}}} and
\protect{\lowercase{\ref{se:nonequilibrium}}}}

\label{se:compare} 
\begin{figure}[ht]
\hspace{0.0\hsize} \epsfxsize=0.99\hsize \epsffile{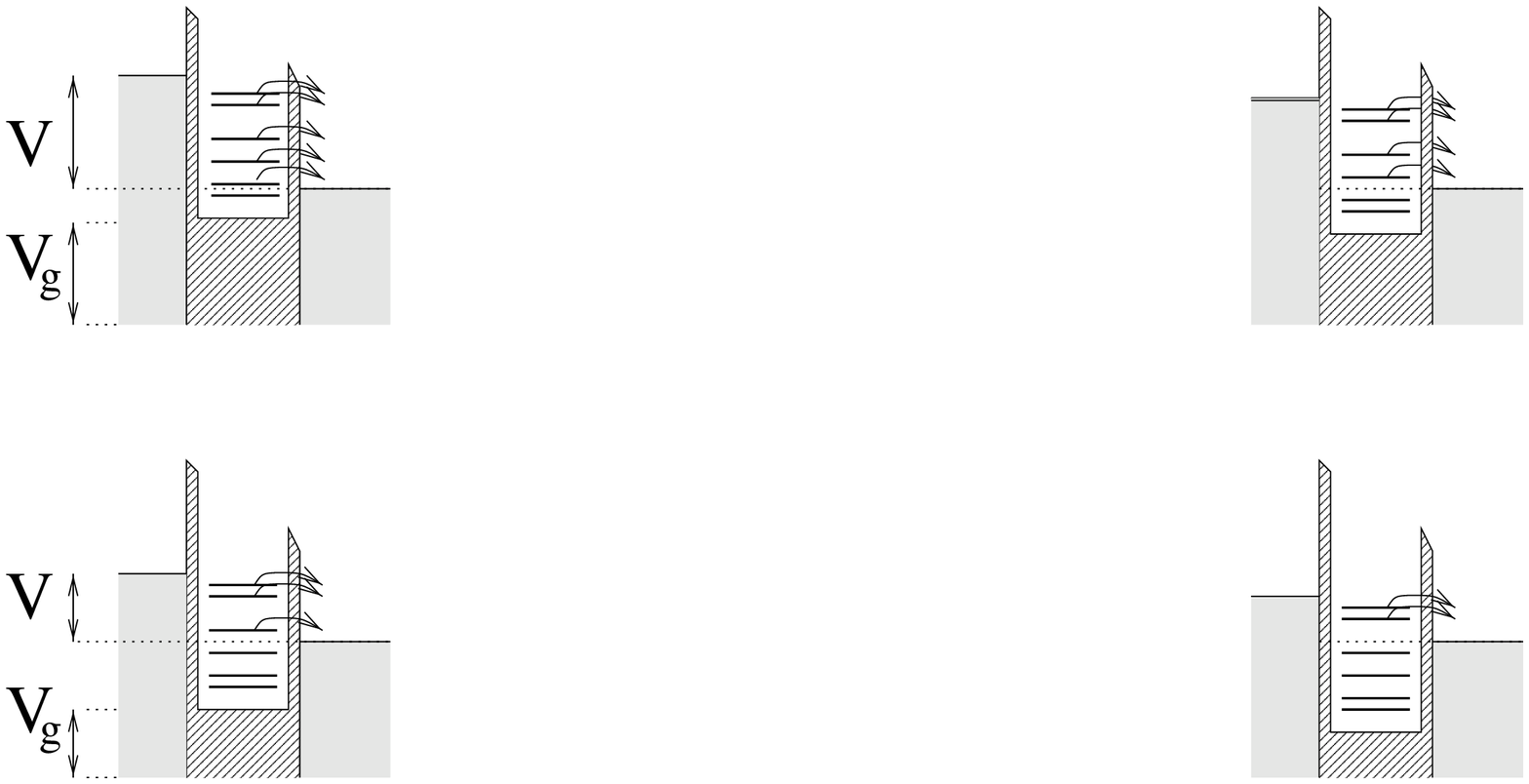}
\vspace{-0.5\hsize}

\hspace{0.21\hsize} \epsfysize=0.295\hsize
\epsffile{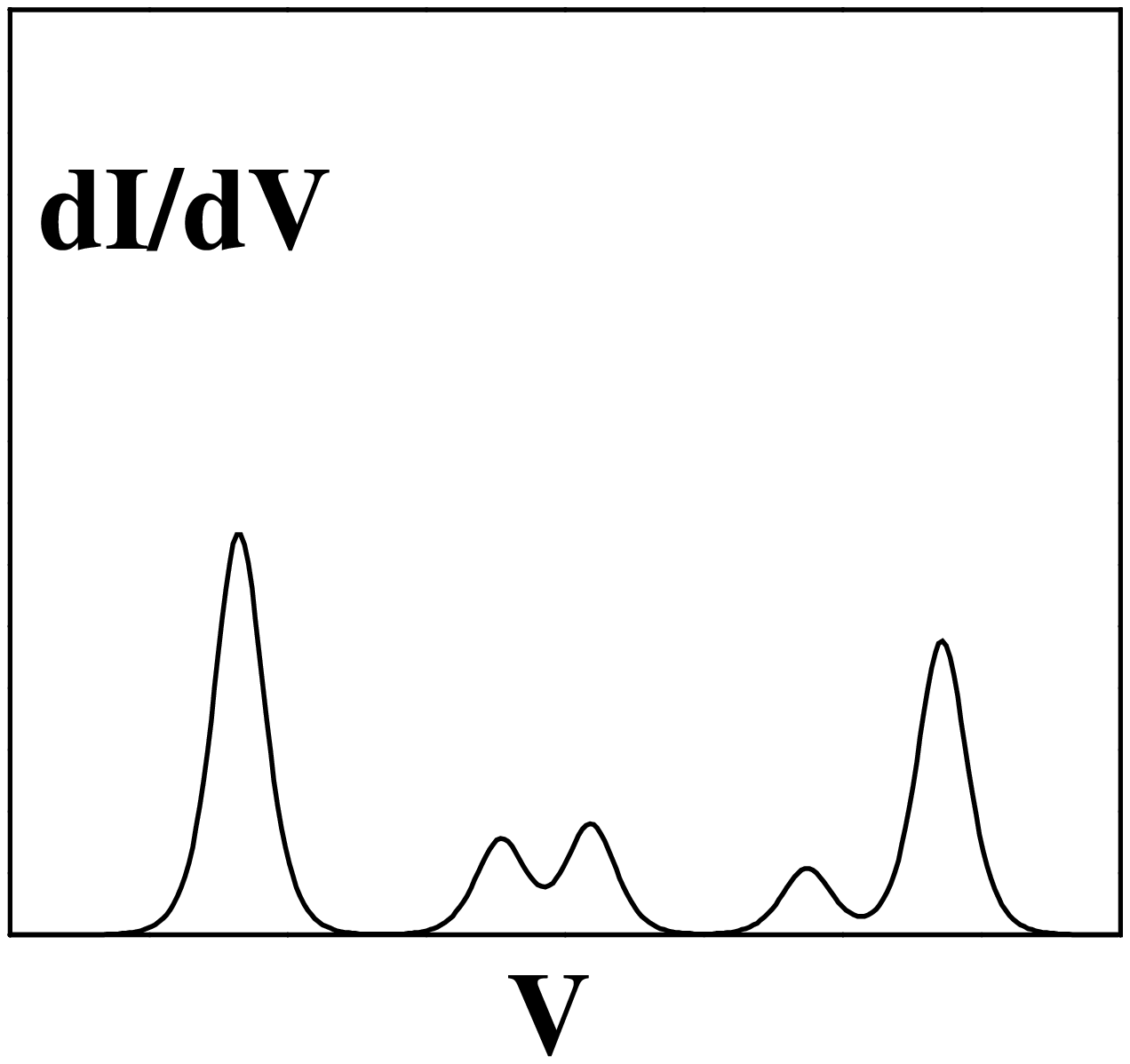} \hspace{-0.12\hsize}
\epsfysize=0.295\hsize \epsffile{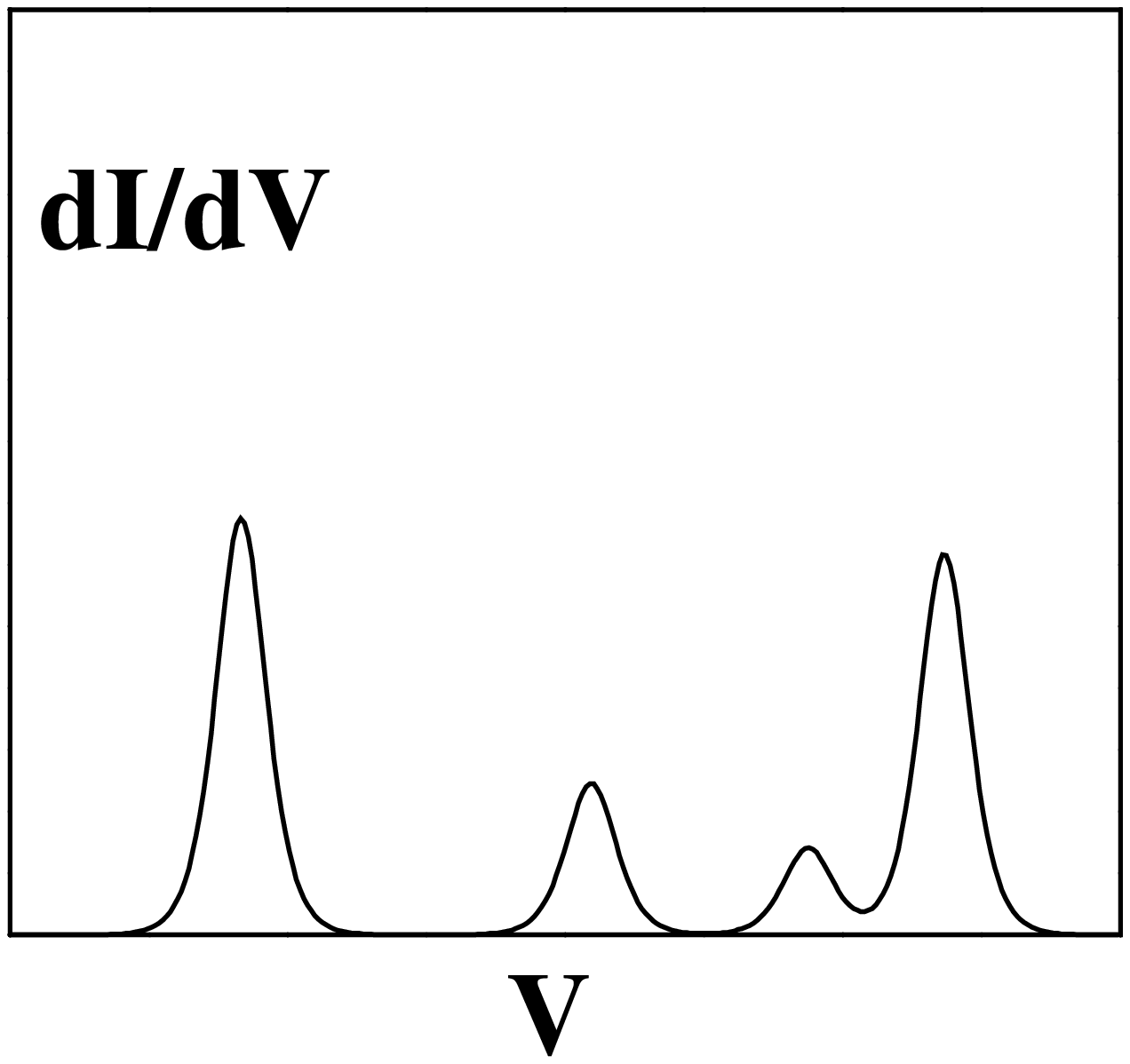}
\vspace{-0.04\hsize}

\hspace{0.21\hsize} \epsfysize=0.295\hsize
\epsffile{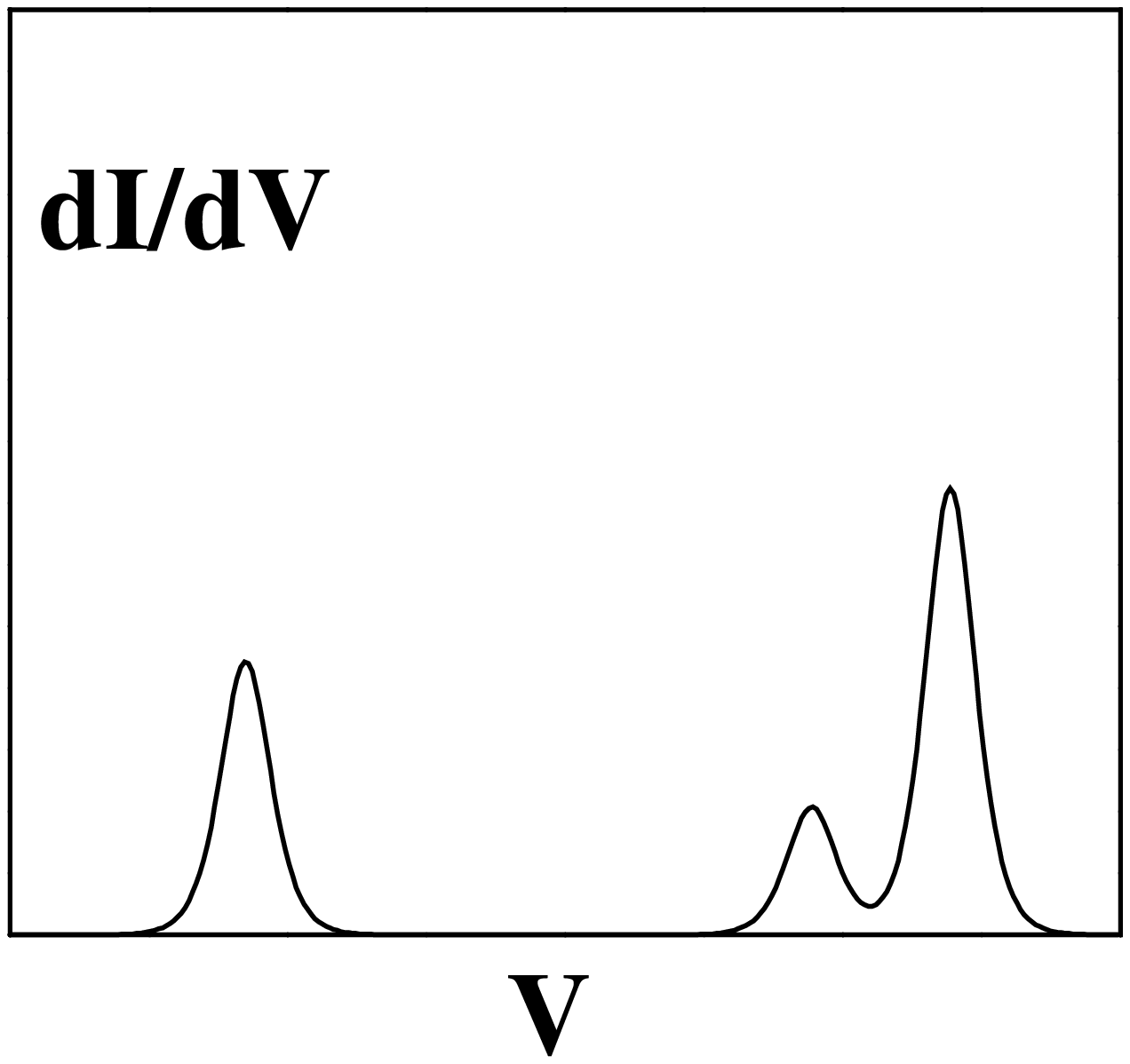} \hspace{-0.12\hsize}
\epsfysize=0.295\hsize \epsffile{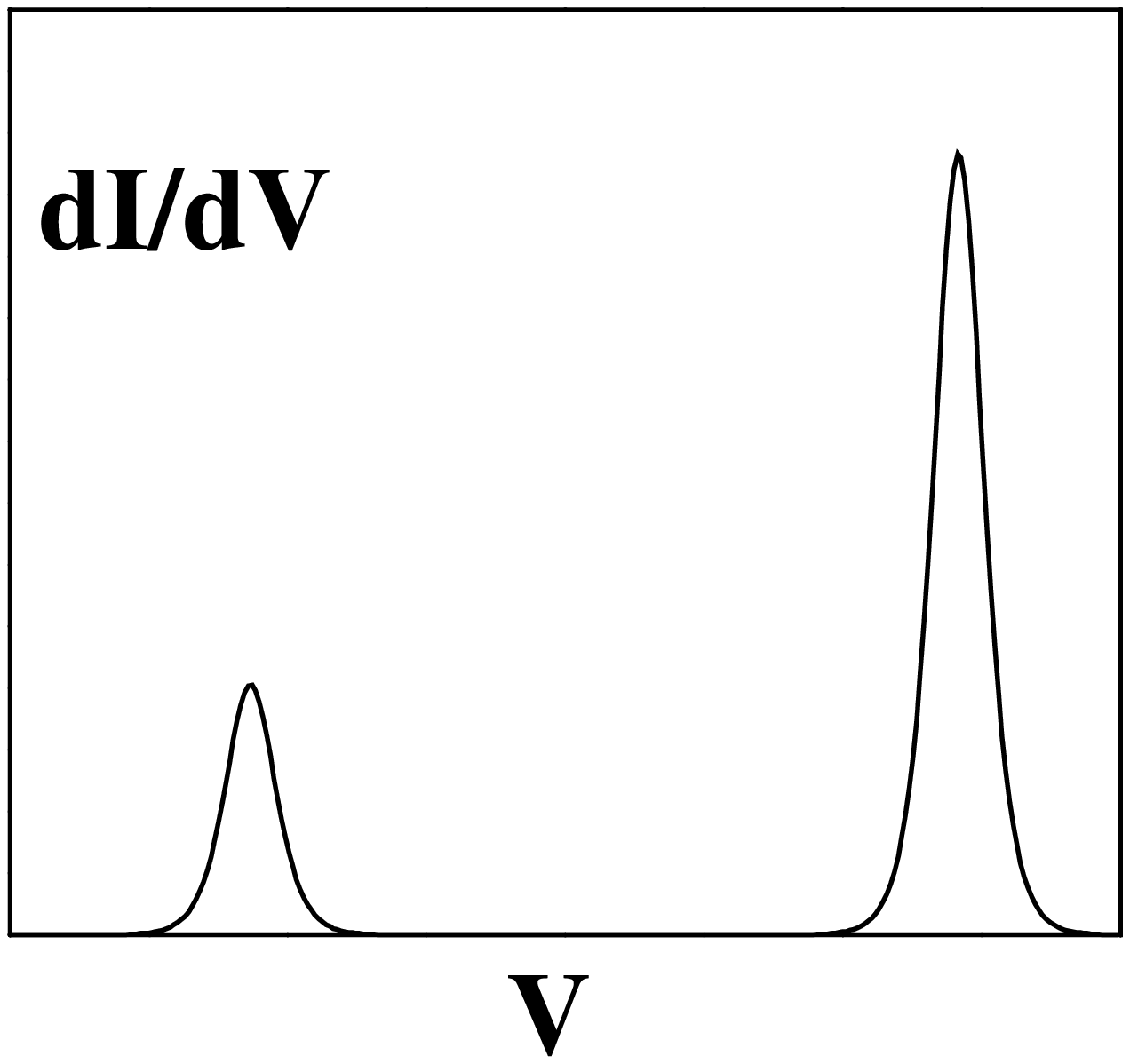}
\vspace{-0.06\hsize}
 \caption{\label{fig:vgate} When the voltage $V_g$ of a
nearby gate is varied, the number of excited $N$-particle states
that can be populated by current flow is also changed. The four
panels show how the peaks in the differential conductance
disappear one by one when $V_g$ is tuned in the direction of the
charge-degeneracy point (at which current flow happens at zero
bias voltage). As seen in the figure, the (minimal) bias voltage,
$V$, for nonzero current flow decreases as $V_g$ approaches the
charge-degeneracy point. In the four panels shown, the number of
peaks decreases from five (upper left) to two (lower right panel).
The single-particle levels are for the $N+1$-particle system; the
arrows indicate from which levels electrons can escape to the
right reservoir. The $dI/dV$ graphs were calculated using the rate
equations of Refs.~\protect\cite{Beenakker,Averin}, with randomly
chosen values for the interaction matrix elements that determine
the dependence of transition energies on the precise population of
the single-particle levels in the grain, see
Subsection~\protect\ref{se:fluctuations}. The typical distance
between the peaks is of order $\Delta/\sqrt{g}$, where $g$ is the
dimensionless conductance of the metal grain. }
\end{figure}

The main difference between the two
explanations offered here is that the former entails many
different transitions between states very close to the $N$ and
$N+1$-particle ground states, whereas the latter makes use of, in
principle, the same transition between different pairs of states
that are highly excited above the ground state. Thus, one may
distinguish between the two scenarios, when the electrostatic
potential of the grain can be changed by the voltage $V_g$ on a
nearby gate: Fine tuning of $V_g$ affects the threshold bias
voltage $V_{\rm th}$, and hence the number of possible
nonequilibrium configurations. Hence, if the fine structure of the
first resonance is due to nonequilibrium processes, the peaks will
disappear one by one when $V_g$ is tuned to the charge degeneracy
point. This is illustrated in Fig.~\ref{fig:vgate}. In this
figure, we have simulated the differential conductance from the
rate equations of Ref.~\cite{Beenakker,Averin}, for the case where
all relaxation of excited states inside the grain occurs due to
the coupling to the leads. The four panels show the differential
conductance for four different values of the gate voltage, where
the multiplet consists of $5$, $4$, $3$, and $2$ peaks. The closer
the gate voltage is to a charge degeneracy point, the fewer
nonequilibrium peaks can be observed. On the other hand, if the
fine structure is due to a degenerate ground state, no highly
excited states are involved, and the number of peaks in the
multiplet is insensitive to $V_g$. Of course, both explanations
(nonequilibrium processes and an almost degenerate ground state)
can apply at the same time. In that case, multiple peaks will
disappear at the same time, when $V_g$ is tuned to the charge
degeneracy point. An alternative way to distinguish between the
two scenarios is if the parity of the number of electrons $N$ can
be changed by a gate voltage. Nonequilibrium processes cannot
explain a fine structure of the first resonance if $N$ is odd.

\section{Conclusions}

In this article, we have reviewed several effects related to the
electrons' spin in the presence of spin-orbit coupling and/or
electron--electron interaction, in small quantum dots with
relatively large number of electrons.

At low temperatures, in the absence of spin-orbit coupling, and in
the absence of an applied Zeeman field, our system can be
described by an effective Hamiltonian of the form (\ref{eq:heff}).
In the  limit where the number of electrons in a chaotic dot is
large, the effective Hamiltonian contains three interaction
parameters, in the direct, exchange and Cooper channels. In
Appendices~\ref{ap:LP} and~\ref{ap:cooper} we estimate these
parameters for realistic systems. We have used use them in
Sec.~\ref{se:interactions} to calculate (i) the probability for
the dot to have a non-zero total spin in its ground state and (ii)
the distribution of the Coulomb blockade peak spacings. The theory
for the latter describes many features of the experimental
observations, and is qualitatively much better than what one would
have obtained if one ignored the exchange interaction.  However,
the simple effective  model does not do well in describing the low
energy tail of the distribution, and it does not account for the
large differences in the data obtained by different experimental
groups.

In Sec.~\ref{se:SO}, we reviewed some  recent theoretical studies
on the effect of spin-orbit coupling in a quantum dot or metal
particle. In Subsection~\ref{se:gtens}  we discussed the relation
between the splitting of the ground state Kramers degeneracy (in
the case of a strong spin-orbit coupling and a weak magnetic
field) and an effective $g$-tensor. The joint probability
distribution of the eigenvalues of the tensor was presented.
Recent experiments \cite{SOC:Petta01} that measured the
distribution of the $g$-tensor in particles of Cu, Ag, and Au,
found good agreement with many aspects of the theoretical
predictions.

The combined effects of spin-orbit coupling and electron-electron
interactions were discussed in Subsection~\ref{se:so+int}. It was
argued that strong spin-orbit coupling will tend to inhibit the
appearance of effects due to electron-electron interactions.

In Subsection~\ref{se:GaAs}, we reviewed how  the peculiar form of
the spin-orbit coupling for a two-dimensional electron system in a
GaAs heterostructure of quantum well leads to a strong suppression
of spin-orbit effects when the electrons are confined in a small
quantum dot. We explained how a magnetic field, parallel to a
quantum dot in a AlGaAs 2DEG, enhances the weak spin-orbit effects
in these dots. This observation  may be used to tune the strength
of spin-orbit coupling in quantum dots and may explain recent
observations on fluctuations in the conductance through such dots.

We also discussed, in Subsection~\ref{se:mutiplets} possible
explanations, based on non equilibrium phenomena, and an almost
degenerate ground states due to spin-orbit coupling and
electron--electron interaction, for the observations
\cite{WF:Davidovic00} of a multiplet splitting of the lowest
resonance in the tunneling conductance through a gold
nanoparticle.

A few recent developments in the young field of spin and
interaction effects in small quantum systems have been examined in
this article. At present, there is no quantitative theory that can
explain all the experimental observations in this area. Current
theories describe well several aspects of small dots (for example
the $g$-tensor eigenvalue distribution), but in others aspects,
such as Coulomb blockade peak spacing, the agreement between
theory and experiment is far from being satisfactory.

%% %% %% %%%%%%%%%%%%%%%%%%%%%%%%%%%%%%%%%%%%%%%%%%%%%%%%%%%%% %%
\section{Acknowledgments}

It is our pleasure to thank our collaborators J. N. H. J. Cremers,
Ady Stern, J.~A.~Folk and C.~M~Marcus. Special thank are due to
S.~R.~Patel and C.~M.~Marcus for allowing us using their data in
this publication and for enlightening conversations. We also thank
M.~Tinkham and D.~Davidovic for helpful discussions, and
S.~L\"{u}scher, T.~Heinzel and K.~Ensslin for sending us their
data. This work was supported at Harvard by NSF grants DMR-9809363
and DMR-9981283, at Cornell by NSF grant DMR-0086509 and by the
Sloan Foundation, and at  Weizmann by the German-Israeli Project
Corporation DIP-c7.1.

\appendix

\section{Derivation of the effective Hamiltonian
[Eq.~(\mbox{\lowercase{\protect\ref{eq:heff}}})] from the toy
model with contact interaction
[Eq.~(\mbox{\lowercase{\protect\ref{eq:ham}}})]}

\label{ap:toy}

In order to analyze the ground state energy for the toy model
Hamiltonian~(\ref{eq:ham}), the electron-electron interaction is
separated into mean and fluctuations,
\begin{eqnarray}
  uM \sum_{m}
      n_{m\uparrow}  n_{m\downarrow},
  &=&
  uM \sum_{m} \left( \langle  n_{m\uparrow}
             \rangle
              n_{m\downarrow}
             +
              n_{m\uparrow}
             \langle
              n_{m\downarrow}
             \rangle \right)
  \nonumber \\ && \mbox{}
  + uM \sum_{m} \delta n_{m\uparrow} \delta  n_{m\downarrow}
  \nonumber \\ && \mbox{}
  - uM \sum_{m} \langle  n_{m\uparrow}
             \rangle \langle
              n_{m\downarrow}
             \rangle,
  \label{eq:HF}
\end{eqnarray}
where $\delta  n_{ms} =  n_{ms} - \langle n_{ms} \rangle$ and the
average occupation $\langle  n_{m,\uparrow} \rangle = \langle
n_{m,\downarrow}\rangle$ is calculated using the self-consistent
(Hartree-Fock) Hamiltonian
\begin{equation}
  {\mathcal H}^{\rm HF} = \sum_{n,m,s}
     c^{\dagger}_{n,s} {\mathcal H}_0(n,m)
       c^{\vphantom{\dagger}}_{m,s}
    +
  uM \sum_{m} \left( \langle  n_{m\uparrow}
             \rangle
              n_{m\downarrow}
             +
              n_{m\uparrow}
             \langle
              n_{m\downarrow}
             \rangle \right).
  \label{eq:HHF}
\end{equation}
For technical convenience, we define the occupancies $\langle
n_{ms}\rangle$ in a reference state with $2N - 2K$ electrons and
zero spin, where $K$ is a number of order unity, chosen such that
all relevant particle-hole excitations have energy less than $K
\Delta$. For a disordered metal grain, we may assume that the
eigenvectors and eigenvalues $\varepsilon_{\mu}^{\rm HF}$ of
${\mathcal H}^{\rm HF}$ are distributed like those of a random
matrix (with the possible exception of the spacing of two
eigenvalues closest to the Fermi level, see
Refs.~\cite{WF:Brouwer99,WF:Levit99}; a reference state with $2N -
2K$ electrons, rather than with $2N$ electrons is chosen, to
ensure that only eigenvalues of ${\mathcal H}^{\rm HF}$ above the
Fermi level are needed). The last term on the r.h.s.\ of
Eq.~(\ref{eq:HF}) is a constant shift of the energy and can be
omitted.

We then construct a state with $2N$ (or $2N+1$) electrons by
adding $2K$ (or $2K+1$) electrons to the reference state, and find
an effective Hamiltonian for low-lying particle-hole excitations
using the remaining third term on the r.h.s.\ of Eq.~(\ref{eq:HF})
as a perturbation. The derivation of this effective Hamiltonian
proceeds along the lines sketched in Sec.\
\ref{sec:GroundStateSpinDistribution} and
Ref.~\cite{WF:Kurland00}. In the limit $M \to \infty$, the
effective Hamiltonian has the form (\ref{eq:heff}), where to
lowest nontrivial order in $u$ one has $\varepsilon_{\mu} =
\varepsilon_{\mu}^{\rm HF}$ and $-J_{\rm s} = J_{\rm c} = u$.

When studying the effective interaction amplitudes perturbatively
in $u$, one finds that virtual particle-hole excitations involving
states with energies $\varepsilon$ far away from $E_{\rm F}$
($\Delta<\varepsilon-E_{\rm F} < E_{\rm Th}$) contribute to the
$O(u^2)$ term and to higher order terms. This leads to an
effective renormaliziation of the interaction constants $J_{\rm
s}$ and $J_{\rm c}$, and of the spacing between the levels
$\varepsilon_{\mu}$. For example,
\begin{eqnarray}
  J_{\rm s}(u) &=& {u} +
  {u^2} {1 \over M} \left[
  \int^{E_{\rm F}} d\varepsilon d\varepsilon'
  {\nu(\varepsilon) \nu(\varepsilon') \over
  \varepsilon + \varepsilon'}
% \right. \nonumber \\ && \left. \mbox{}
  - \int_{E_{\rm F}} d\varepsilon d\varepsilon'
  {\nu(\varepsilon) \nu(\varepsilon') \over
  \varepsilon + \varepsilon'} \right]
  + {\mathcal O}(u^3), \label{eq:Jsu} \\
  \varepsilon_{\mu}(u) &=&
  \varepsilon_{\mu}^{\rm HF}
   - u^2 \varepsilon_{\mu}^{\rm HF} {1 \over M}
  \left[\int^{E_{\rm F}} d\varepsilon d\varepsilon'
  \int_{E_{\rm F}} d\varepsilon''
  {\nu(\varepsilon) \nu(\varepsilon') \nu(\varepsilon'')
  \over
  (\varepsilon + \varepsilon' - \varepsilon'')^2}
  \right. \nonumber \\ && \left. \mbox{} +
  \int^{E_{\rm F}} d\varepsilon
  \int_{E_{\rm F}} d\varepsilon' d\varepsilon''
  {\nu(\varepsilon) \nu(\varepsilon') \nu(\varepsilon'')
  \over
  (\varepsilon - \varepsilon' - \varepsilon'')^2} \right]
% \nonumber \\ && \mbox{}
  + {\mathcal O}(u^3),
\end{eqnarray}
 where $\nu(\varepsilon)$ is the density of states for the
Hamiltonian ${\mathcal H}^{\rm HF}$, and $J_{\rm c}$ is
renormalized to zero (see appendix~\ref{ap:cooper}).

Beyond the first order in the interaction strength $u$, the
symmetry of the effective Hamiltonian (\ref{eq:heff}), and of the
result (\ref{eq:EGdiff}) that was derived from it, differs from
that of the equivalent expression in Ref.~\cite{WF:Brouwer99},
which was obtained from the toy model (\ref{eq:ham}) using the
selfconsistent Hartree-Fock approximation. The reason of this
difference is that, in higher orders of perturbation theory, the
selfconsistent Hartree-Fock approximation neglects certain
contributions to the ground state energy. (For example, the first
correction to $J_{\rm s}$ is of second order in $u$ [second term
in Eq.~(\ref{eq:Jsu})], and not, as in the Hartree-Fock
approximation, of third order\cite{WF:Brouwer99}.)

When all contributions are taken into account, the symmetry of
Eq.~(\ref{eq:EGdiff}) and the form of the effective Hamiltonian
(\ref{eq:heff}) is preserved to all orders in $u$.

\section{The relation between the parameter
$\mathbf{\lowercase{\lambda}=-J_{\rm \lowercase{s}}/\Delta }$ and
$\mathbf{r_{\rm s}}$  } \label{ap:LP}

Landau-Fermi-liquid theory expresses various properties of the
system in terms of the coefficients $f_{p \sigma p \sigma'}$.
Using the notations of Ref.~\cite{SS:Pines66} we find:
\begin{equation}
\frac{C}{C_b}=\frac{m^\star}{m_b}=1+\frac{F_1^s}{d};\;\;
\frac{\chi_P}{\chi_{Pb}}=\frac{m^\star}{m_b} \frac{1}{1+F_0^a},
%\frac{\kappa}{\kappa_b}=\frac{m^\star}{m_b}\frac{1}{1+F_0^s},
\end{equation}
where $C$ is the specific heat, $\chi_P$ the Pauli
susceptibility. The quantities with the suffix $-_b$ include band
effects, but do not include electron--electron interaction
corrections. The latter are encompassed in the $F$-coefficients of
the Landau Fermi liquid theory. The letter $d$ denotes the
dimension of the system.
For ballistic systems there are no anomalous renormalizations of
the Fermi-Liquid coefficients and we have:
\begin{equation}
\lambda=-J_{\rm s}/\Delta=-F_0^{\rm a} =
1-\frac{\chi_P}{\chi_{Pb}}\frac{m^\star}{m_b} =
1-\frac{\chi_{Pb}}{\chi_{P}}\frac{C}{C_b}.
\end{equation}
Thus, in principle the ratio of the specific heat and the
susceptibility gives the desired interaction parameter $J_{\rm
s}$.

\subsection{Three dimensions}

There are various ways to calculate theoretically the Landau $F$
parameters, using different approximations for the
electron--electron interaction. They varied from a simple static
RPA approximation to more complicated approaches like density
functional theory. Ref.~\cite{WF:Singwi81} reviews the subject.

The relevant parameter to describe the strength of the interaction
effects is~$r_{\rm s}$, the ratio of the typical potential energy
to the kinetic energy of electrons. In metals it is defined by:
\begin{equation}
\label{def:rs3} \frac{4 \pi}{3} r_{\rm s}^3 a_B^3= \frac{1}{n},
\;\; a_B = \frac{e^2 m_b} {\hbar^2 \epsilon 4 \pi \epsilon_0}
\end{equation}
where $n$ is the density of electrons and $a_B$ is the Bohr radius
in the metal. Some values for $r_{\rm s}$ in metals are giving in
Ref.~\cite{WF:Hedin69} (page 74) and in Ref.~\cite{SS:Ashcroft87}.

 For $0<r_{\rm s}< 5$ the effective mass\cite{WF:Hedin69} ranges between $0.96<
 m^\star/m_b <1.06$ where for small $r_{\rm s}$ ($<3$), $m^\star < m_b$, and for larger $r_{\rm s}$ ($>3$), $m^\star > m_b$.
 (See table VII on page 103 in \cite{WF:Hedin69}.)

Using the approximation of Rice \cite{RFS:Rice65} for the
effective mass and for the susceptibility one can roughly
approximate
\begin{equation}
\label{eq:lambda3d1} \lambda_{\rm Rice}(r_{\rm s}) \sim (3+r_{\rm
s})/25,\; {\rm for}\; 1< r_{\rm s} <5.
\end{equation}

Another approximation for the susceptibility is given in
\cite{WF:Singwi81} see page 256. Assuming that the effective mass
is renormalized as in the Rice approach (i.e. not very significant
renormalization) we find that:
\begin{equation}
\label{eq:lambda3d2} \lambda_{\rm Singwi}(r_{\rm s}) \sim
(2+r_{\rm s})/16,\; {\rm for}\; 1< r_{\rm s} <5.
\end{equation}

For small $r_{\rm s} \sim 1$ the difference between the
estimations is only 15 \% while for $r_{\rm s} \sim 5$ it close to
30 \%.
%
%In any case $-J_{\rm s} < 0.5$ even for $r_{\rm s}=5$ with the optimistic
%assumption that we use saying that $m^\star$ did not renormalize
%significantly the susceptibility.
%
The second estimate reproduces quite well experimental
measurements of $\lambda$ in a wide range of metals. The parameter
$\lambda$ is determined by various experimental methods such as
electron spin resonance, spin wave, Knight shift and total
susceptibility. [See p.~256 of Ref.~\cite{WF:Singwi81} and
reference therein for further details.]

 Using Eqs.~(\ref{eq:lambda3d1}) and (\ref{eq:lambda3d2}) we can estimate the
parameter $J_{\rm s}$ in different materials; however the
estimates are rough and should be taken with a grain of salt.
Typical values for metallic elements are given in the table.
\begin{center}\
\begin{table}[ht]\hskip 2cm
\begin{tabular}{c|cccccccccccccccc}
\hline
&&&&&&&&\\
                 & Li & Na & K & Rb & Cs & Cu & Ag & Au \\
\hline
&&&&&&&&\\
$r_{\rm s}$ & 3.25& 3.93& 4.86& 5.20& 5.62& 2.67& 3.02& 3.01 \\
$\lambda_{\rm Rice}(r_{\rm s})$ & 0.25& 0.28& 0.31& 0.33& 0.34& 0.23& 0.24& 0.24 \\
$\lambda_{\rm Singwi }(r_{\rm s})$ & 0.33& 0.37& 0.43& 0.45& 0.48& 0.29& 0.31& 0.31 \\
\hline \hline
&&&&&&&&\\
                 & Be & Mg & Ca & Sr & Ba & Nb & Fe & Pb \\
\hline
&&&&&&&&\\
$r_{\rm s}$ & 1.87& 2.66& 3.27& 3.57& 3.71& 3.07& 2.12& 2.30 \\
$\lambda_{\rm Rice }(r_{\rm s})$ & 0.19& 0.23& 0.25& 0.26& 0.27& 0.24& 0.20& 0.21 \\
$\lambda_{\rm Singwi}(r_{\rm s})$ & 0.24& 0.29& 0.33& 0.35& 0.36&
0.32& 0.26& 0.27
\end{tabular}
\vskip 0.5cm
 \caption{\vskip -0.95cm \hskip 2cm\label{table1} Estimations for $\lambda=-J_{\rm s}/\Delta$ in
selected metals. \vskip 0.5cm}
\end{table}
\end{center}

\subsection{Two dimensions}

Most of the calculations for the Landau-Fermi-liquid parameters in
two dimensional systems were performed for Silicon MOSFET. For a
review see Ref.~\cite{WF:Singwi81} (especially page 257) and
Ref.~\cite{WF:Ando82} (pages 454 and 468).

We note that as we sweep an external magnetic field, perpendicular
to the sample area, the spin susceptibility oscillates because the
difference in the occupations of Landau levels with spins up and
down. This effect makes the comparison between theory and
experiment complicated. We will not be interested in such
anomalously large exchange enhancement.

In case of silicon MOSFET we should include also the valley
degeneracy, and the difference in the dielectric constants of Si
and SiO that causes the dielectric function be space dependent.
The screening from the metallic electrodes may influence the
results as well.

For GaAs/AlGaAs heterostructures the first two complications are
absent. It appears that due to the absence of valley degeneracy in
GaAs/AlGaAs the parameter $J_{\rm s}$ should be larger than in the
case of the Si MOSFET. Therefore, GA/AlGaAs might be more
appropriate to the study of spin configurations and there
dependance on interaction constants.

A static random phase approximation for GaAs gives
\cite{WF:Janak69}
\begin{equation}
\label{eq:lambdastatic} \lambda=\frac{m^\star}{m_b} G(\frac{r_{\rm
s}}{\sqrt{2}});\;\;\; G(x)= \left\{
\begin{array}{ccc}
\frac{x}{\pi} \frac{\arccos{\! \rm h}(1/x)}{\sqrt{1-x^2}} & \mbox{
for } & x \le 1
\\
\frac{x}{\pi} \frac{\arccos(1/x)}{\sqrt{x^2-1}} & \mbox{ for } & x
> 1
\end{array} \right.
\end{equation}

where in two dimensions
\begin{equation}
\label{def:rs2} r_{\rm s}= \frac{e^2 m_b}{\hbar^2\sqrt{\pi
n}}\frac{1}{\epsilon 4 \pi \epsilon_0}
=\frac{5.45*10^5}{\sqrt{n(cm^2)}}.
\end{equation}
In the last equality we take $\epsilon=12.9,\; m_b=0.067 m_e$ and
$m_e$ is the free electron mass.

It can be verified that
$$
 G(x) \stackrel{x \rightarrow \infty}{\longrightarrow} 1/2 \;\;\;\;{\rm
 and}\;\;\;\;
G(x) \stackrel{x \rightarrow 0}{\longrightarrow} (x/\pi)\log(2/x).
$$
 The factor 1/2 for large $x$ is due to the spin degeneracy and
appears because both spins participate in the screening in the RPA
approximation.(In case of a MOSFET the factor 1/2 is substituted
by 1/4 due to the valley degeneracy.)

The same static RPA approximation gives for the effective mass:
\begin{equation}
\label{eq:meff} m_b/m^\star = 1-(\sqrt{2}/\pi) r_{\rm s}+r_{\rm
s}^2/2+(1-r_{\rm s}^2)G(r_{\rm s}/\sqrt{2}).
\end{equation}
Numerically, in this approximation $0.95 < m^\star/m_b < 1$. In
other words within the static RPA approximation the mass
renormalization is not very significant.

\begin{figure}[ht]
\vglue -0.2cm \epsfxsize=0.5\hsize
\epsffile{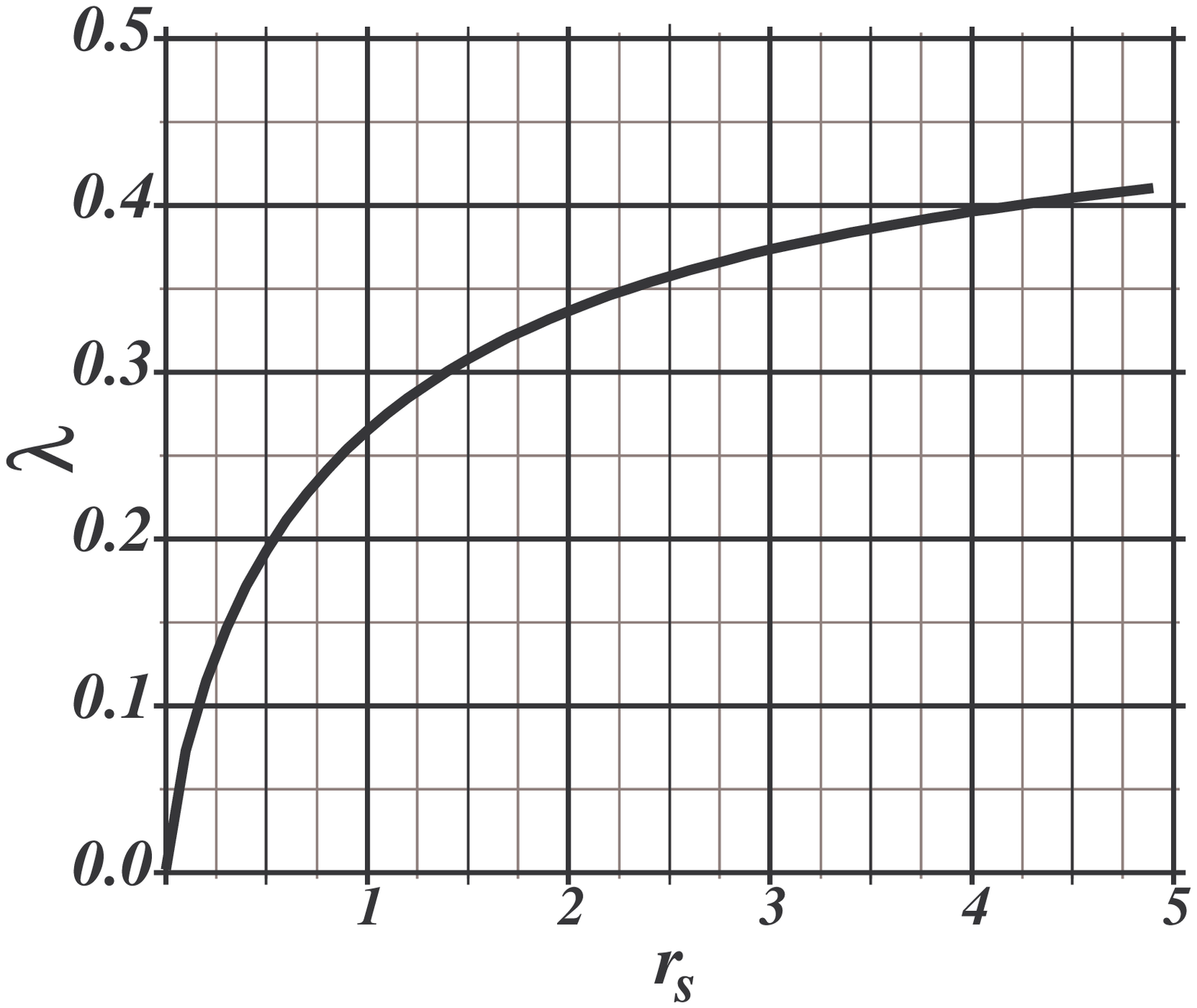}\vspace{-0.425\hsize}
\vglue 0cm \hspace{0.52\hsize} \epsfxsize=0.524\hsize
\epsffile{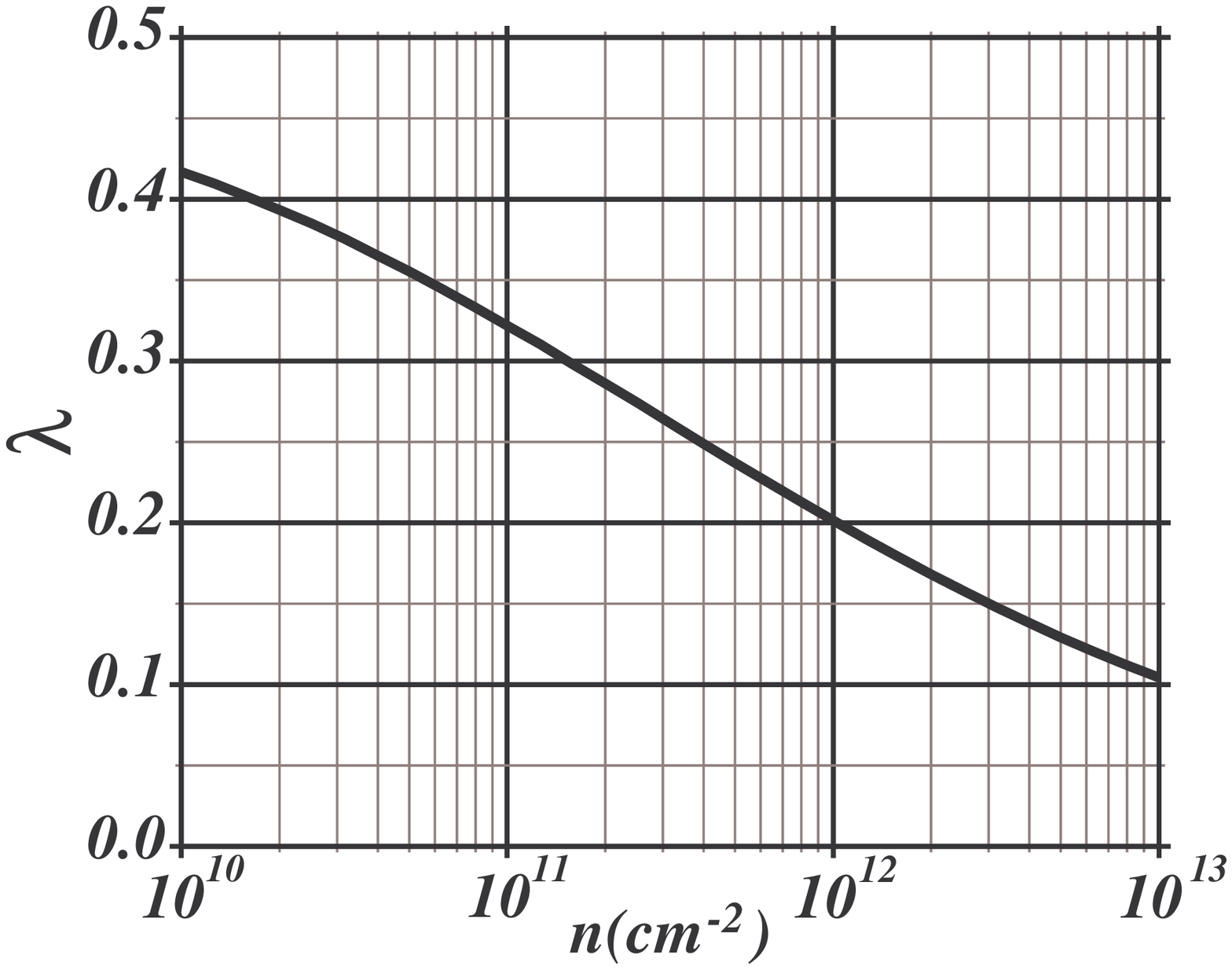}\vspace{0cm}
 \caption{ \label{fg:lambda1} $\lambda$ as a
function of  the ratio of the typical potential energy to the
kinetic energy of electrons $r_{\rm s}$, and as a function of the
electron density $n$, for GaAs/AlGaAs heterostructures in the
static RPA approximation.}
\end{figure}

Using this approximation we plot $-J_{\rm s}$ as a function of
$r_{\rm s}$ and $n$ in Fig.~\ref{fg:lambda1}. For example, $
\lambda(n = 0.7*10^{11} cm^{-2}) \cong \lambda(r_{\rm s} \cong 2)
\sim 0.34$,

\section{Renormalization of the interaction in the Cooper channel}
\label{ap:cooper}

In Sec.~\ref{se:Heff} we described how to integrate out the
interaction between electrons at high frequency in the RG sense.
The interaction in the Cooper channel deserves a special
consideration, as we will see below it reduces substantially when
the temperature decreases.

To see how it works in practice we look at the Dyson equation for
the vertex part of the interaction in the Cooper channel (for a
precise  definition of the vertex part see Ref.~\cite{RFS:AGD63}
Sec. 33.3). Since the divergencies in the Cooper channel are
logarithmic we can write the Dyson equation, for $T$ larger than
the inverse of the elastic mean free time $\tau$, in a RG
form\cite{DS:Finkelstein90}:
\begin{equation}
\label{eq:RG} {dJ_{\rm c}(T)}/{d l} =
-J^2_c(l)/\Delta,\;\;l=\log(E_{\rm F}/{T}), \;\; T> 1/\tau.
\end{equation}
Integration of this equation, from $E_{\rm F}$ to $T$ gives
\begin{equation}
J_{\rm c}(T) = \frac{J_{\rm c}(E_{\rm F})}{1+{(J_{\rm c} (E_{\rm
F})/\Delta)\log{(E_{\rm F}/T)}}}. \label{eq:coopera}
\end{equation}
This logarithmic suppression of the interaction in the Cooper
channel was first discussed in Ref.~\cite{SC:Morel62} and is known
as the Tolmachev-Anderson-Morel $\log$ or
pseudo-electron-potential $\log$.

In quasi-two-dimensional samples, for $T< 1/\tau$ the RG equation
(\ref{eq:RG}) is modified to \cite{DS:Finkelstein90}:
\begin{equation}
\label{eq:RG1} {dJ_{\rm c}(T)}/{d l} = \Delta/(g
\pi)-J^2_c(T)/\Delta,\;\;
 1/\tau> T > E_{\rm Th}.
\end{equation}
We have neglected here the effects of the diffusive motion on the
other channels. The presence of the term $1/(\pi g)$ slows down
the logarithmic reduction in the Cooper channel, and physically
describes the enhancement of the interaction between the electrons
in the Cooper channel due to their diffusive motion. In case of
quasi-one-dimensional systems the full Dyson equation should be
solved\cite{SCW:Oreg99}.

Finally, the process of integration of the motion at high
frequencies reaches the Thouless energy, and we find the effective
Hamiltonian (\ref{eq:heff}). We will analyze the Cooper channel
for energies below the Thouless energy using the contact model
(\ref{eq:ham}). In principle, the behavior of the Cooper channel
can be solved exactly by the method of Richardson
\cite{SC:VonDelft01}. But, to understand qualitatively the
reduction of the interaction in the Cooper channel, at energy
below $E_{\rm Th}$ it is sufficient to solve the Dyson equation
for the interaction matrix element $\left<\alpha \bar \alpha
\right| \hat u \left| \bar \alpha \alpha \right>$ in the Cooper
channel. Formally this equation is:
\begin{equation}
 \left<\alpha \bar \alpha \right| \hat u \left| \alpha \bar \alpha \right> =
 \left<\alpha \bar \alpha \right| \hat u^0 \left| \alpha \bar \alpha \right> - \sum_{\nu
  \not = \alpha} \frac{\left<\alpha \bar \alpha \right| \hat u^0 \left| \nu \bar \nu
  \right>\left<\nu \bar \nu \right| \hat u \left| \alpha \bar \alpha
  \right>}{\left|\varepsilon_a-\varepsilon_\nu \right|}, \nonumber
\end{equation}
 with $ \hat u^0 = uM c^{\dagger}_{n,\uparrow} c^{\dagger}_{n,\downarrow}
c^{\vphantom{\dagger}}_{n,\downarrow}
c^{\vphantom{\dagger}}_{n,\uparrow} \delta_{nm}$, $ \hat u =
c^{\dagger}_{n,\uparrow} c^{\dagger}_{n,\downarrow} u(n,m)
c^{\vphantom{\dagger}}_{m,\downarrow}
c^{\vphantom{\dagger}}_{m,\uparrow}$, the operator $ \
\psi^\dagger_{\alpha \uparrow(\downarrow)} = \sum_{n=1}^M
\phi_{\alpha \uparrow(\downarrow)} (n) c^\dagger _{n,
\uparrow(\downarrow)}$ where the functions $\phi_\mu(n)$ are real
eigenfunctions of a random matrix with real elements, $n$ runs on
the sites of the random systems, $M$ is the total number of sites,
and $\left| \alpha \bar \alpha \right>=  \psi^\dagger_{ \alpha
\downarrow} \psi^\dagger_{\alpha \uparrow} \left|0 \right>$. Using
the anti-commutation relations of $c_{n,s}$ operators we find an
equation for the unknown amplitudes $u(n,m)$
\begin{eqnarray}
 \sum_{n m} \phi^2_\alpha(n) u(n,m) \phi^2_\alpha(m)= \nonumber \qquad\qquad\qquad \\
 \sum_{n m} \phi^2_\alpha(n) (u M \delta_{nm}) \phi^2_\alpha(m)+ \qquad\qquad\qquad \\
 \sum_{n,m,l, k,\nu \not = \alpha} \frac{\phi^2_\alpha(n) (u M \delta_{nl})
  \phi^2_\nu(l)\phi^2_\nu(k) u(k,m ) \psi_\alpha^2(m)} {\varepsilon_\alpha-\varepsilon_\nu}. \nonumber
\end{eqnarray}
 Now with the relation $\left<\phi^2_\mu(l)\phi^2_\mu(k) \right>=
1/M^2(1+2\delta_{lk})$ (that is valid in the limit $M\rightarrow
\infty$) we find, comparing the elements in the series that sum
over $n$ and $m$
\begin{equation}
\label{eq:unm} u(n,m)=
   uM\delta_{nm}-\lambda \frac{\log M}{M} \left(\sum_l{u(l,m)}+2u(n,m)\right),
\end{equation}
where the logarithmic factor appears from the summation over the
energies $\nu$. The term $\propto 2 u(n,m)$ on the left hand side
can be neglected as it is small [by a factor $(\log M)/M$]
compared to the one on the right hand side. A summation over $n$
gives now
$$\sum_l u(l,m) = \frac{uM}{1+\lambda \log M}$$
substituting in (\ref{eq:unm}) and taking the limit $M \gg 1$ we
find:
\begin{equation}
\label{eq:unmsol} u(n,m)=\frac{M \delta_{nm}+\lambda \log M
(M\delta_{nm}-1)} {1+\lambda \log M},
\end{equation}
where $\lambda=u/\Delta$ and $\Delta$ is the average level spacing
in the dot. Hence:
\begin{equation}
\left<\alpha \bar \alpha \right| \hat u \left| \alpha \bar \alpha
\right>= \frac{1}{M^2}\sum_{m,n} (1+2\delta_{nm}) u(n,m) = 2
u+\frac{u}{1+\lambda \log M}.
\end{equation}
The logarithmic factor reduces the term that is $\propto
\frac{1}{M^2}\sum u(n,m)$ and does not involve the $\delta_{nm}$
factor. This term correspond to the Cooper channel as it involves
contraction of two wave functions, associated with two creation
(or two annihilation) operators. Thus, we find that from the
Thouless energy up to energies of the order of the level spacing,
similarly to the pseudo-electron-potential $\log$ [see after
Eq.~({\ref{eq:coopera})], there is an additional logarithmic
suppression of the interaction in the Cooper channel. For that
reason we took $J_{\rm c}=0$ in the analysis of the ground state
spin distributions in Sec.~\ref{sec:GroundStateSpinDistribution}
and the Coulomb blockade peak spacing in Sec.~\ref{se:CB}.

\end{document}